\documentclass[a4paper, 12pt]{article}

\usepackage[utf8]{inputenc}
\usepackage{amsmath, amssymb}
\usepackage{graphicx}
\usepackage{caption}
\usepackage{booktabs}
\usepackage{geometry}
\usepackage{lipsum} 
\usepackage{natbib}
\usepackage{hyperref}
\usepackage{lscape}
\usepackage{cleveref} 

\hypersetup{
    colorlinks=true,
    linkcolor=blue,
    filecolor=magenta,      
    urlcolor=cyan,
    citecolor=blue,         
    pdftitle={Research Paper},
    pdfpagemode=FullScreen,
}

\title{Forecasting Thai inflation from univariate Bayesian regression}
\author{
    Paponpat Taveeapiradeecharoen\thanks{PhD, paponpat.tav@mfu.ac.th} \and Popkarn Arwatchanakarn
    \thanks{Assistant Professor in Economics, popkarn.arw@mfu.ac.th}
}
\date{\today}

\begin{document}

\maketitle

\begin{abstract}
This study investigates the forecasting performance of Bayesian shrinkage priors in predicting Thai inflation in a univariate setup, with a particular interest in comparing those more advance shrinkage prior to a likelihood dominated/noninformative prior. Our forecasting exercises are evaluated using Root Mean Squared Error (RMSE), Quantile-Weighted Continuous Ranked Probability Scores (qwCRPS), and Log Predictive Likelihood (LPL). The empirical results reveal several interesting findings: SV-augmented models consistently underperform compared to their non-SV counterparts, particularly in large predictor settings. Notably, HS, DL and LASSO in large-sized model setting without SV exhibit superior performance across multiple horizons. This indicates that a broader range of predictors captures economic dynamics more effectively than modeling time-varying volatility. Furthermore, while left-tail risks (deflationary pressures) are well-controlled by advanced priors (HS, HS+, and DL), right-tail risks (inflationary surges) remain challenging to forecast accurately. The results underscore the trade-off between model complexity and forecast accuracy, with simpler models delivering more reliable predictions in both normal and crisis periods (e.g., the COVID-19 pandemic). This study contributes to the literature by highlighting the limitations of SV models in high-dimensional environments and advocating for a balanced approach that combines advanced shrinkage techniques with broad predictor coverage. These insights are crucial for policymakers and researchers aiming to enhance the precision of inflation forecasts in emerging economies.
\end{abstract}

\section{Introduction}

There are quite few in terms of investigating on how accurate on Bayesian regression model forecast for Thai inflation. Despite that multiple literature are worth discussing. To begin with

For frequentist approach, on the other hand, we have ..

Inflation forecasting remains a critical area of research in macroeconomics, with various methodologies being employed to enhance predictive accuracy. Studies on inflation dynamics have largely focused on assessing the role of domestic and global factors, evaluating different econometric models, and exploring the effectiveness of monetary policy frameworks. Here we categorize the inflation literature as followed: First we would like to introduce about inflation volatility \textbf{the Global vs. Domestic Drivers of Inflation}: Research on inflation in Thailand and other economies frequently examines whether inflation is primarily driven by domestic monetary conditions or external factors. For example, \cite{manopimoke2018thai} highlights the increasing influence of global variables, such as oil prices and world output gaps, on Thailand’s inflation after 2001. However, it also underscores the continued relevance of domestic monetary policy in the long run. Similarly, \cite{hossain2020time} finds that inflation in Thailand and Indonesia is more sensitive to external shocks than domestic shocks, reinforcing the idea that inflation forecasting models must account for both internal and external economic forces. Our second category is \textbf{Inflation Forecasting Models}: Several econometric techniques have been applied to inflation forecasting, ranging from traditional time-series models to modern machine learning and Bayesian methods. Structural Vector Autoregression (SVAR) models have been widely used to capture inflation’s response to macroeconomic shocks. As an illustration from \cite{gongsiang2021inflation}, where the research utilizes quantile regressions and skewed-t distributions to assess the risks surrounding future inflation, particularly in times of financial instability. However, these approaches often assume static relationships between variables, which may not be valid in the presence of structural economic shifts. 

Dynamic models such as Dynamic Stochastic General Equilibrium (DSGE) models have also been applied to inflation forecasting, with \cite{luangaram2022exploring} using Bayesian estimation to demonstrate that the Bank of Thailand (BOT) adjusts interest rates in response to exchange rate movements to stabilize inflation. The use of Bayesian stochastic volatility models, as seen in \cite{koirala2023inflation}, has gained traction in capturing the time-varying nature of inflation uncertainty across G20 countries. However, country-specific models are necessary to address unique policy and economic conditions.

Bayesian econometric techniques have gained prominence in inflation forecasting due to their ability to incorporate prior knowledge, address parameter uncertainty, and adapt to structural economic shifts. Traditional Bayesian Vector Autoregressions (BVARs) \cite{litterman1986forecasting,koop2010bayesian,giannone2015prior} often rely on standard shrinkage priors like the Normal-Inverse Wishart or Minnesota prior. However, recent advances in hierarchical shrinkage priors—such as the Horseshoe prior \cite{carvalho2010horseshoe,makalic2015simple}, Dirichlet-Laplace prior \cite{bhattacharya2015dirichlet}, Horseshoe+ \cite{bhadra2017horseshoe+}, Lasso \cite{bhadra2019lasso}, Ridge \cite{hoerl1970ridge,bedoui2020bayesian}, and Spike-and-Slab—offer \cite{ishwaran2005spikemicroarray,hernandez2013generalized,bai2021spike} significant improvements in forecasting accuracy, particularly for economies like Thailand, where inflation is influenced by complex and time-varying factors.

Advanced priors play a crucial role in Thai inflation forecasting by capturing sparse but relevant predictors. The Dirichlet-Laplace prior and Horseshoe prior offer adaptive shrinkage, aggressively pushing irrelevant predictors toward zero while preserving the influence of key inflation drivers such as oil prices, exchange rates, and global financial conditions, see for instance \cite{koop2014new,cross2019macroeconomic}. Given Thailand's small open economy, where external shocks often dominate domestic inflation dynamics, these priors ensure that only the most relevant variables are included in the forecasting model.  

Structural breaks and regime shifts have significantly influenced Thai inflation, \cite{manopimoke2018thai}, and among others \cite{hossain2016inflation,hossain2020time}, including events such as the post-1997 financial crisis, the transition to inflation targeting in 2000, and supply-chain disruptions caused by COVID-19. Horseshoe+ and Spike-and-Slab priors could potentially be particularly effective in handling such shifts, as they dynamically adjust the model's level of sparsity, allowing it to capture policy-driven or external changes in inflation dynamics.  

Forecast accuracy during periods of high volatility is another key challenge in inflation forecasting, especially during sudden inflation surges like those experienced during the 2008 global financial crisis or the recent post-pandemic inflation shocks, among others \cite{lenza2020estimate,huber2020nowcasting,hauzenberger2021gaussian,clark2021tail,clark2024forecasting}. The Horseshoe+ prior, an extension of the original Horseshoe, offers stronger local shrinkage properties while preventing excessive shrinkage of critical predictors. This makes it especially effective in macroeconomic forecasting, where it is essential to account for abrupt but significant changes without distorting key signals.  

Furthermore, Thai inflation forecasting often necessitates incorporating a large number of potential drivers, see for example \cite{hossain2020time}, including monetary aggregates, global inflation, and commodity prices. This increases the risk of overfitting in traditional Bayesian Vector Autoregression (BVAR) models. As demonstration in \cite{cross2019macroeconomic}, macroeconomic variables rather somewhere between dense and sparse, not entirely sparsed. Therefore it is logical for us to implement advanced priors such as the Dirichlet-Laplace and Horseshoe+ mitigate this issue by controlling overfitting while still capturing inflation-relevant signals, making them particularly well-suited for high-dimensional forecasting models.

Having said all of the above, this study aims to evaluate the forecasting performance of various Bayesian shrinkage priors for Thai inflation. We compare multiple prior specifications—Ridge, Lasso, Dirichlet-Laplace, Horseshoe, and Horseshoe+, using a noninformative prior as the benchmark. By systematically analyzing their predictive capabilities, we assess the advantages of each prior in improving inflation forecasts. To ensure robustness, we examine two different predictor settings within each prior specification: Moderate-sized models with approximately 20 predictors, which represent a standard macroeconomic forecasting approach.
High-dimensional models with up to 56 predictors, which capture a broader set of economic variables, allowing us to analyze how these priors handle increasing model complexity.
Furthermore, we adopt a direct forecasting approach for predicting inflation at different horizons. Multistep direct forecasts offer key advantages over iterative methods by reducing the propagation of model specification errors and enhancing stability, particularly in volatile macroeconomic conditions. This approach ensures that the predictive accuracy reflects the intrinsic properties of the priors rather than cumulative estimation errors.

Another critical aspect of our study is the evaluation of prior performance during crisis periods, specifically the 2008 global financial crisis and the COVID-19 pandemic. These events introduced severe turbulence in inflation volatility, and we aim to investigate how different priors respond to such instability. By analyzing inflation forecasts during these periods, we assess each prior's ability to adapt to abrupt structural shifts and economic shocks.

\section{Methodology}
\subsection*{Bayesian Linear Regression}
In this research, we employ a univariate Bayesian linear regression model to forecast Thailand's inflation. The general form of a Bayesian linear regression model is given by:
\begin{equation}
y = X\beta + \epsilon,
\end{equation}

where, \( y \) is the \( n \times 1 \) vector of the response variable, representing the inflation rate, \( X \) is the \( n \times p \) design matrix or covariates of predictor variables. \( \beta \) is the \( p \times 1 \) vector of regression coefficients. Finally the vector of error terms \( \epsilon \) with \( n \times 1 \) dimension, which is assumed to follow a normal distribution \( \epsilon \sim N(0, \sigma^2 I) \).

The Bayesian approach to linear regression involves specifying a likelihood function for the observed data and a prior distribution for the parameters. The likelihood is quite straightforward which is given by:
\begin{equation}
y \mid X, \beta, \sigma^2 \sim N(X\beta, \sigma^2 I).
\end{equation}

The posterior distribution is obtained by combining the likelihood with a prior distribution on the regression coefficients \( \beta \) and the error variance \( \sigma^2 \) using Bayes' theorem, \cite{koop2003bayesian}:
\begin{equation}
p(\beta, \sigma^2 \mid y, X) \propto p(y \mid X, \beta, \sigma^2) p(\beta \mid \sigma^2) p(\sigma^2).
\end{equation}

The choice of prior distributions is crucial in Bayesian regression, influencing both the model's regularization properties and its out-of-sample forecasting performance. In this study, we investigate the forecasting performance of six priors. We first begin by introducing the noninformative (mildly informative) which has been a work horse for Bayesian linear regression for a while. The prior is mildly in a sense that it is assumed to have zero mean prior with unit standard deviation, allowing likelihood to dominate mostly while also moderately shrink element in the vector of regression coefficients centered around zero. This prior will be our benchmark, see \cite{marcellino2006comparison}. Next is Ridge \cite{hoerl1970ridge}, , Lasso \cite{tibshirani1996regression}, Horseshoe \cite{carvalho2010horseshoe,makalic2015simple}, Horseshoe Plus, and Spike-and-Slab, \cite{mitchell1988bayesian}.

\subsection*{Noninformative Prior}
A noninformative (or mildly informative) prior imposes minimal assumptions on the coefficients, typically using a diffuse normal prior:
\begin{equation}
\beta \mid \sigma^2 \sim N(0, c \cdot \sigma^2 I),
\end{equation}

where \( c \) is a large constant, reflecting vague prior knowledge. In our work we set this number into $10^4$. The reason we use this as our benchmark lies in its benefit, as mentioned before, it allows the data to primarily influence the posterior, making it suitable for exploratory analysis. Additionally it avoids the risk of introducing strong biases from prior beliefs. Such prior is useful when little prior knowledge is available or when a more objective analysis is desired which is suitable for macroeconomic data such as Thai consumer price index here. Despite that there are cons that worth mentioning such that it can lead to over-fitting and poor predictive performance in high-dimensional settings. However we only use this prior as a the AR(2)-Bayesian-regression therefore we can ignore such high-dimensional issues.

This prior is considered in this study to provide a benchmark against more informative shrinkage priors, enabling a fair comparison of forecasting accuracy. Also we would like to investigate if adding more macroeconomic variables will improve the out-of-sample forecasts. That is those additional macroeconomic variables once added into the Bayesian regression model, is it sparse or we are facing the sparse illusion like the previous literature that mentioned, \cite{cross2019macroeconomic}. Such paper investigate the global-local shrinkage prior then augment into the Bayesian vector autoregression (BVAR) model while adding stochastic volatility. They found out that the macroeconomic variables are not sparse but rather dense, leading to the one of the most used prior such as Minnesota also known as Litterman prior, see \cite{litterman1986forecasting}, performs so well in terms of US macroeconomic forecasts. 

The competitive priors that we will be using are described in the following sub-sections.
\subsection*{Ridge Prior}
The first of penalized regression priors we would like to briefly introduce is the Ridge prior, \cite{hoerl1970ridge}, which imposes a Gaussian prior on the regression coefficients. Although it is quite old but it is still being used for comparative reasons, see \cite{polson2014bayesian,bedoui2020bayesian}, among others:
\begin{equation}
\beta \mid \sigma^2 \sim N(0, \lambda^{-1} \sigma^2 I),
\end{equation}

where \( \lambda \) is called a global shrinkage hyperparameter that controls the overall magnitude of the regression coefficients but applies the same level of shrinkage to all coefficients uniformly. Unlike local shrinkage parameters in Horseshoe prior (which allow different levels of shrinkage for each coefficient), and will be introduced later on, the Ridge prior assumes no sparsity and does not differentiate between important and unimportant predictors. A higher value of \( \lambda \) results in greater shrinkage of the coefficients towards zero, effectively reducing model complexity.

Despite the restriction of the global shrinkage hyperparameter where we have to set in advance before the approximation is executed via Gibbs-sampling method the benefit of this prior are that it mitigates multicollinearity issues by shrinking correlated predictors, and it is computationally efficient and straightforward to implement. The Gibbs-sampling for conditional posterior distribution can be implemented straightforwardly, see \cite{bedoui2020bayesian}. Ridge prior is particularly beneficial in this study because inflation forecasting often involves multicollinear economic predictors, making this prior suitable for enhancing stability and interpretability.

\subsection*{Adaptive Lasso Prior}
The Lasso prior introduces sparsity by assuming a Laplace (double-exponential) prior on the coefficients:
\begin{align}
\begin{split}
\beta_j \mid \sigma^2 &\sim \mathcal{L}(0, \lambda_j),\\
	\lambda_j &= \frac{1}{|\widehat{\beta}_j|^{\gamma}+\epsilon}.
\end{split}
\label{eq:adaptivelasso1}
\end{align}
where $\widehat{\beta}_j$ is an initial estimate (e.g. from OLS, or Ridge). \( \lambda \) is the localized-regularization parameter. This prior induces sparsity by shrinking some coefficients exactly to zero, thus performing variable selection. The Lasso is different to the ridge in a sense that it allows the $\lambda$ to learn with the data within Gibbs loops, ensuring that it will be the optimal value based on each data being used. We will implement both ridge and lasso to test whether those additional macroeconomic variables are significant in terms of out-of-sample forecast Thai inflation since the Lasso prior is ideal when only a few predictors are significant, enabling both regularization and feature selection. To this day such prior is still being developed and extended, see for example, \cite{park2008bayesian} which reformulates Lasso in a Bayesian framework by placing Laplace (double-exponential) priors on regression coefficients, and use hierarchical representation to facilitate Gibbs sampling, providing posterior distributions for parameter uncertainty. Or Adaptive Lasso from \cite{leng2014bayesian}. We also add $\epsilon$ in \cref{eq:adaptivelasso1} for numerical stability.

Finally $\gamma>0$ controls the degree of adaptiveness in the Adaptive Lasso prior. It modifies the shrinkage $\lambda_j$ for each coefficient based on its magnitude from an initial OLS/other estimate.

If $\gamma = 1$, standard adaptive lasso \cite{zou2006adaptive}, if $\gamma > 1 $, more aggressive shrinkage for small coefficients, and if $\gamma < 1$, weaker shrinkage (closer to Ridge regression).

\subsection*{Spike-and-Slab Prior}
The Spike-and-Slab prior, \cite{george1993variable,ishwaran2005spike}, is a mixture prior that explicitly performs variable selection:
\begin{align}
\begin{split}
    \beta_j \mid \gamma_j &\sim (1-\gamma_j) \delta_0 + \gamma_j \mathcal{N}(0, \sigma^2), \\
    \gamma_j \mid \pi &\sim \text{Bernoulli}(\pi), \\
    \pi &\sim \text{Beta}(a, b), \\
    \sigma^2 &\sim \mathcal{IG}(\alpha, \beta),
\end{split}
\end{align}
where \( \gamma_j \) is a binary indicator (0 or 1), controlling whether the coefficient is exactly zero (spike) or normally distributed (slab), in other words, included/excluded from the regression model. $\delta_0$ is a point mass at zero (the spike), or Dirac delta function, representing the spike at zero. Next is $\pi$ which is the inclusion probability (common across all predictors). $\sim \text{Bernoulli}(\pi)$ is Bernoulli distribution for the indicator variable. $\text{Beta}(a,b)$, is Beta prior for the inclusion probability, allowing hierarchical modeling of sparsity. Finally $\sim{\mathcal{IG}}(\alpha,\beta)$, is Inverse-Gamma prior for the variance of the slab, allowing uncertainty in the size of non-zero coefficients.

\subsection*{Horseshoe Prior}
The horseshoe prior, \cite{carvalho2010horseshoe} is a continuous, global-local shrinkage prior designed to handle sparse signals. It is characterized by heavy tails and a peak near zero, making it highly adaptive in shrinking noise while preserving large signals. Horseshoe prior is considered to be adaptive shrinkage, where it aggressively shrinks small coefficients towards zero while allowing large coefficients to remain mostly unaffected, making it highly effective for sparse models. Unlike Lasso, it does not require a hard threshold for feature selection. Additionally its heavy tail prevents over-shrinkage of large signals, preserving important features which may contain a good source of out-of-sample predictive power. Finally it is a continuous shrinkage especially relative to the spike-and-slab, where it avoids the computational complexity of discrete mixture models. The Horseshoe prior can be recognised as followed:
\begin{align}
\begin{split}
\beta_j \mid \lambda_j, \tau &\sim \mathcal{N}(0, \lambda_j^2 \tau^2\sigma^2),\\
	\sigma^2 &\sim\sigma^{-2}d\sigma^2,\\
		\lambda_j&\sim{\mathcal{C}^{+}}(0,1),\\
			\tau&\sim{\mathcal{C}^{+}}(0,1).
\end{split}
\end{align}
Or equivalently,
\begin{align}
\begin{split}
\lambda_j^{-1} \mid \nu_j^{-1} &\sim \mathcal{G}\left(\frac{1}{2}, \nu_j^{-1}\right), \\
\tau^{-2}\mid \xi^{-1} &\sim \mathcal{G}\left(\frac{1}{2}, \xi^{-1}\right), \\
\nu_1^{-1}, \ldots, \nu_j^{-1}, \xi^{-1} &\sim \mathcal{G}\left(\frac{1}{2}, 1\right).
\end{split}
\end{align}
where \( \lambda_j \) is a local shrinkage parameter for each coefficient, and \( \tau \) is a global shrinkage parameter. $X\sim{\mathcal{G}}(a,b)$ is Gamma distribution with shape and rate as $a$, and $b$, respectively. Such that formula one can obtain conditional posterior distribution via Gibbs method straightforwardly, see \cite{makalic2015simple}. For interested readers for such prior for high-dimensional issues, are referred to \cite{bhadra2019lasso}.

The Horseshoe prior is slightly computationally more intensive than the previous introduced priors due to the hierarchical structure of the prior but those computational time is barely noticeable.

\subsection*{Horseshoe Plus Prior}
An extension of the Horseshoe prior, the Horseshoe Plus prior, \cite{bhadra2017horseshoe+}, enhances flexibility by introducing an additional layer of local shrinkage:
\begin{align}
\begin{split}
    \beta_j \mid \lambda_j, \tau, \nu_j &\sim \mathcal{N}(0, \lambda_j^2 \tau^2), \\
    \lambda_j|\phi_j,\xi &\sim \mathcal{C}^+(0, \phi_j \xi), \\
    \phi_j &\sim \mathcal{C}^+(0, 1), \\
    \xi &\sim \mathcal{C}^+(0, 1), \\
    \tau &\sim \mathcal{C}^+(0, 1),
\end{split}
\end{align}
Or equivalently
\begin{align}
\begin{split}
    \beta_j \mid \lambda_j, \tau, \nu_j &\sim \mathcal{N}(0, \lambda_j^2 \tau^2), \\
    \lambda_j^{-2}\mid\phi_j, \xi &\sim \mathcal{G}\left( \frac{1}{2}, \frac{\phi_j \xi}{2} \right), \\
    \phi_j^{-2} &\sim \mathcal{G}\left( \frac{1}{2}, \frac{1}{2} \right), \\
    \xi^{-2} &\sim \mathcal{G}\left( \frac{1}{2}, \frac{1}{2} \right), \\
    \tau^{-2} &\sim \mathcal{G}\left( \frac{1}{2}, \frac{1}{2} \right).
\end{split}
\end{align}
The key feature of Horseshoe plus prior relative to its original Horseshoe is that it has double layer of local shrinkage, meaning that, $\lambda_j$ and $\phi_j$ jointly control the local shrinkage, enhancing adaptability. In addition there is extra global flexibility i.e. $\xi$, and $\tau$ together govern the global shrinkage, offering a more flexible tail behavior. One worth to mention is that it has a better tail robustness when compared to the standard Horseshoe, Horseshoe+ provides even heavier tails, preventing over-shrinkage of large coefficients.

We chose this prior to investigate its adaptability and enhanced sparsity control in the context of Thai inflation forecasting.

\section{Scoring Matrices}
\subsection*{Root Mean Squared Forecast Error}
RMSE is commonly used in economic forecasting, time series analysis, and machine learning due to its interpretability and sensitivity to large errors. A lower RMSE value indicates a better-fitting model. However, it treats all forecast errors symmetrically and does not differentiate between overprediction and underprediction. The Root Mean Square Error is defined as:
\begin{equation}
    \text{RMSE} = \sqrt{\frac{1}{N} \sum_{t=1}^{N} (y_t - \hat{y}_t)^2}
\end{equation}

where $y_t$ represents the actual observed values, $\hat{y}_t$ denotes the predicted values, and $N$ is the total number of observations.

The relative of RMSE can be computed as followed : 
\[
\text{relative-RMSE}(\text{prior}_1, \text{benchmark}) = \frac{\text{RMSE}(\text{prior}_1)}{\text{RMSE}(\text{benchmark})}
\]

\subsection*{Quantile Weighted Continuous Ranked Probability Score}
The Quantile Weighted Continuous Ranked Probability Score (qwCRPS) is an extension of the CRPS that emphasizes certain regions of the predictive density, such as the upper tail, making it particularly useful for economic applications where the cost of underestimating inflation is higher than overestimating it. qwCRPS is also useful in scenarios where the accuracy of certain quantiles is more critical than overall performance. This is particularly relevant in inflation forecasting, where underestimating inflation can have severe economic implications. Compared to RMSE, qwCRPS provides a \textbf{full density evaluation} rather than just a pointwise error measure. The qwCRPS is defined as, \citet{gneiting2007strictly,gneiting2011comparing}:

Given a realized value \( y \) and a predictive density \( f \), let \( F \) represent the cumulative distribution function (CDF) associated with \( f \). The quantile function of \( F \) at a given probability level \( q \in (0,1) \) is denoted by \( F^{-1}(q) \). The CRPS can be equivalently defined in three distinct mathematical forms:  
\begin{align}
\mbox{CRPS}(f,y) &= \mathbb{E}_F|Y - y| - \frac{1}{2} \mathbb{E}_F|Y - Y'|, \label{eq:CRPS1}\\
	&= \int_{-\infty}^{\infty} \left( F(z) - \mathbb{I}\{y \leq z\} \right)^2 dz, \label{eq:CRPS2}\\
		&= 2\int_{0}^{1} \left( \mathbb{I}\{y < F^{-1}(q)\} - q\right)\left(F^{-1}(q) - y\right) dq. \label{eq:CRPS3}
\end{align}  
In these definitions, \( Y \) and \( Y' \) are independent random variables drawn from the distribution \( F \), while \( \mathbb{I}(\cdot) \) represents the indicator function. Expression \cref{eq:CRPS2} corresponds to the standard CRPS, often called the Brier score, while \cref{eq:CRPS3} represents an alternative form based on quantiles. 

A particularly useful variation of CRPS, referred to as the threshold-weighted CRPS, extends \cref{eq:CRPS2} by incorporating a weighting function \( u(z) \), which allows the evaluation to emphasize specific regions of the predictive distribution:  
\begin{equation}
\label{eq:thresholdWeightedCRPS}
\mbox{S}(f,y) = \int_{-\infty}^{\infty} \left( F(z) - \mathbb{I}\{y \leq z\} \right)^2 u(z) dz.
\end{equation}  
As noted by \cite{gneiting2007probabilistic}, weighting functions like \( u(z) \) enable forecasters to focus on areas of particular interest, such as extreme events or specific quantiles of the distribution. Although threshold decomposition CRPS does not deviate significantly from the Brier score in most practical applications, it remains an insightful extension worth considering.

Beyond uniform CRPS and its threshold-weighted extension, an additional approach is to evaluate forecast accuracy through quantile-weighted CRPS. This measure assesses the accuracy of forecasts at specific quantiles and is given by:  
\begin{equation}
\label{eq:quantileScores}
QS_{\pi}(q,y) = (y - q)(\pi - \mathbb{I}\{y \leq q\}),
\end{equation}  
where \( q \) represents the predicted quantile, and \( \pi \) is the corresponding probability level. The overall weighted CRPS can then be approximated using a discrete sum:  
\begin{equation}
\label{eq:approxQWscores}
S(f,y) = \frac{1}{J-1} \sum_{j=1}^{J-1} v(\pi_j) QS_{\pi_j}(q,y),
\end{equation}  
where \( \pi_j = j/J \) and different weight functions \( v(\pi) \) allow the evaluation to emphasize specific parts of the forecast distribution. In this study, we implement 19 quantiles, ranging from \( 0.05 \) to \( 0.95 \) in increments of \( 0.05 \). Different weighting schemes, as summarized in Table \cref{tab:QuantileWeights}, allow the assessment to prioritize uniform weighting, center regions, tails, or even specific directional emphasis such as the left or right tail. These quantile-weighted scores provide a more nuanced evaluation of density forecasts, making them particularly useful in applications where extreme values or tail risks are of primary concern.

\begin{table}[ht!]
\centering
\resizebox{.45\columnwidth}{!}{%
\begin{tabular}{ll}
\hline
\textbf{Emphasis}   & \textbf{Quantile Weight} \\\hline
uniform          & $v(\pi) = 1$             \\
centre             & $v(\pi) = \pi(1-\pi)$    \\
tails             & $v(\pi) = (2\pi - 1)^2$  \\
right tail        & $v(\pi) = \pi^2$           \\
left tail         & $v(\pi) = (1 - \pi)^2$  \\\hline
\end{tabular}}
\caption{Quantile weights.}
\label{tab:QuantileWeights}
\end{table}

Several studies have used RMSE and qwCRPS for evaluating forecast performance, particularly in Bayesian and time series econometrics. Notable references include \citet{jore2010combining} analyze density forecasts using weighted scoring rules in macroeconomic forecasting, \citet{clark2011real} investigate density forecast evaluation in inflation modeling, and \citet{carriero2020nowcasting,huber2020nowcasting} among others.

Finally the relative of qwCRPS can be computed as followed : 
\[
\text{relative-qwCRPS}(\text{prior}_1, \text{benchmark}) = \frac{\text{qwCRPS}(\text{prior}_1)}{\text{qwCRPS}(\text{benchmark})}
\]

\section{Forecasting Thai inflation using shrinkage prior Bayesian regression}
\begin{table}[ht!]
\centering
\resizebox{\textwidth}{!}{%
\begin{tabular}{lcccclcccc}
\cline{2-5} \cline{7-10}
               & \multicolumn{4}{c}{\textbf{Full periods}}                          &  & \multicolumn{4}{c}{\textbf{Pandemic Period}}                       \\ \cline{2-10} 
               & \multicolumn{9}{c}{\textbf{UC-SV-iterated}}                                                                                                \\ \cline{2-10} 
               & \textit{h = 1} & \textit{h = 4} & \textit{h = 8} & \textit{h = 12} &  & \textit{h = 1} & \textit{h = 4} & \textit{h = 8} & \textit{h = 12} \\ \cline{2-10} 
UCSV           & 0.89           & 1.36           & 0.98           & 0.88            &  & 0.92           & 1.23           & 0.91           & 0.75            \\ \cline{2-10} 
               & \multicolumn{9}{c}{\textbf{AR(2)}}                                                                                                         \\ \cline{2-10} 
DL             & 0.40           & 0.65           & 1.99           & 1.19            &  & 0.44           & 0.72           & 1.49           & 1.07            \\
HS             & 0.30           & 2.41           & 1.16           & 0.98            &  & 0.37           & 2.15           & 1.06           & 0.92            \\
HS+            & 0.27           & 0.97           & 0.92           & 0.88            &  & 0.35           & 0.97           & 0.87           & 0.79            \\
LASSO          & 1.00           & 0.73           & 0.82           & 1.63            &  & 0.99           & 0.80           & 0.77           & 1.32            \\
Ridge          & 0.40           & 0.64           & 1.97           & 1.14            &  & 0.44           & 0.72           & 1.49           & 1.06            \\
Spike-and-Slab & 0.30           & 2.45           & 1.17           & 0.94            &  & 0.37           & 2.17           & 1.06           & 0.88            \\ \cline{2-10} 
               & \multicolumn{9}{c}{\textbf{Moderately sized models}}                                                                                       \\ \cline{2-10} 
DL             & 0.31           & 0.68           & 1.47           & 1.17            &  & 0.34           & 0.63           & 1.12           & 1.02            \\
HS             & 0.27           & 1.27           & 1.13           & 1.08            &  & 0.30           & 1.05           & 0.93           & 0.90            \\
HS+            & 0.24           & 0.87           & 0.94           & 2.35            &  & 0.25           & 0.85           & 0.85           & 1.73            \\
LASSO          & 0.46           & 0.76           & 3.64           & 1.45            &  & 0.42           & 0.71           & 2.46           & 1.22            \\
Ridge          & 0.35           & 0.73           & 1.49           & 1.18            &  & 0.34           & 0.63           & 1.11           & 1.02            \\
Spike-and-Slab & 0.27           & 1.26           & 1.15           & 0.97            &  & 0.33           & 1.05           & 0.97           & 0.86            \\ \cline{2-10} 
               & \multicolumn{9}{c}{\textbf{Large sized models}}                                                                                            \\ \cline{2-10} 
DL             & 0.21           & 0.91           & 2.65           & 1.79            &  & 0.19           & 0.74           & 1.38           & 0.74            \\
HS             & 0.16           & 1.90           & 1.70           & 1.45            &  & 0.15           & 1.14           & 0.82           & 0.66            \\
HS+            & 0.15           & 1.22           & 1.71           & 6.64            &  & 0.13           & 1.07           & 0.77           & 4.17            \\
LASSO          & 0.44           & 1.05           & 1.54           & 2.69            &  & 0.30           & 1.08           & 0.92           & 1.77            \\
Ridge          & 0.39           & 1.01           & 2.66           & 1.97            &  & 0.38           & 1.19           & 1.81           & 1.62            \\
Spike-and-Slab & 0.16           & 1.83           & 1.82           & 1.37            &  & 0.17           & 1.17           & 0.76           & 0.95            \\ \hline
\end{tabular}%
}
\caption{Relative Root Mean Square Error (RMSE) to the benchmark (Noninformative prior Bayesian AR(2) model) of Different Forecasting Models/Priors.}
\label{tab:rmse_results}
\end{table}


\subsection*{Data and Forecasting Setup}
In this section presents how the out-of-sample forecasts are set, we starts by describe some of the dataset we used, see \cref{subsec:appendixDATA}. Our dataset consists of 340 monthly observations after stationary transformation, spanning from 1996:M1 to 2024:M4. The data covers wide range of economic indicators, to begin with of course the price indices: any variables that related to inflation and price levels. Production and industrial output: i.e, variables related to the production of goods, particularly in the automotive and palm oil industries since these are large amount for Thai economy. We also add government revenue such as tax revenues, and other government crucial incomes. Then sales and trade especially the motor vehicle. Finally labor market and employment such as job vacancies, placement and oversea workers. These categories help in understanding the different aspects of the economy that these variables represent, such as inflation, industrial output, fiscal policy, labor market conditions, and trade activities.

The forecasting models are applied to three different predictor sets: a large-sized model (56 variables)\footnote{We figured that such transformation reduce the seasonal effects and ensure stationarity in the data.}, a moderate-sized model (20 variables), and a small-sized model (AR(2) only). Forecasts are rolled forward beginning with 128 observations, with initial predictions starting from 2006:M9 and extending to 2024:M4. Additionally, we assess forecast performance exclusively during the pandemic period (2019:M12 – 2023:M5).

Multi-step direct forecasting is employed as it prevents the error accumulation common in recursive methods, leading to potentially more accurate long-term predictions \cite{taieb2012recursive}. By training separate models for each forecast horizon, this approach captures complex patterns, improves interpretability, and enhances robustness against anomalies \cite{chevillon2007direct}. Unlike recursive forecasting, direct forecasting allows for parallel computation, improving computational efficiency. However, we also include the widely used Unobserved-Component Stochastic Volatility (UC-SV) model \cite{stock2007has}, which follows an iterated forecasting approach. Prior research \cite{marcellino2006comparison, mccracken2019empirical} suggests that direct multi-step (DMS) and iterated multi-step (IMS) forecasting models perform differently depending on the economic context, with DMS models often excelling for nominal variables, particularly during stable economic periods.

\subsection*{Point accuracy across priors}
In this sub-section the relative root mean square error (RMSE) of various Bayesian priors for forecasting Thailand’s inflation at different horizons ($h = 1, 4, 8, 12$). The Bayesian regression: AR(2) model with a noninformative prior serves as the benchmark. A relative RMSE below 1 indicates superior forecast accuracy compared to the benchmark, whereas values above 1 suggest inferior predictive performance.

We first demonstrate the relative RMSE in \cref{tab:rmse_results}. For the full periods, the UC-SV-iterated model shows strong performance at $h=1$ and $h=12$, with scores of 0.89 and 0.88, respectively, indicating it outperforms the benchmark. However, at $h=4$, its score increases to 1.36, indicating worse performance. During the pandemic period, the UC-SV-iterated model maintains its strong performance at $h=1$ and $h=12$, with scores of 0.92 and 0.75, respectively, while its performance at $h=4$ improves slightly to 1.23.

The AR(2) models, including DL, HS, HS+, LASSO, RIDGE, and Spike-and-Slab, show varying performance across different horizons. At $h=1$, most models perform well, with scores significantly below 1, indicating superior performance compared to the benchmark. For example, the HS+ model has a score of 0.27, and the Spike-and-Slab model has a score of 0.30. However, at longer horizons, such as $h=4$ and $h=8$, the performance of these models deteriorates, with scores often exceeding 1. The HS model, for instance, has a score of 2.41 at $h=4$ during the full periods, indicating much worse performance than the benchmark.

The moderately sized models show a similar pattern. At $h=1$, most models perform well, with scores below 1. The HS+ model, for example, has a score of 0.24 during the full periods. However, at longer horizons, the performance varies. The LASSO model has a score of 3.64 at $h=8$ during the full periods, indicating poor performance, while the Ridge model maintains relatively stable performance across horizons.

The large-sized models also exhibit strong performance at $h=1$, with scores well below 1. The HS+ model, for instance, has a score of 0.15 during the full periods. However, at longer horizons, the performance of these models becomes more inconsistent. The HS+ model has a score of 6.64 at $h=12$ during the full periods, indicating significantly worse performance than the benchmark. During the pandemic period, the large-sized models generally show improved performance at longer horizons, with the HS+ model achieving a score of 4.17 at $h=12$, which, while still high, is better than its performance during the full periods.

Focusing on the large-sized model setup, which incorporates a rich set of up to 56 predictors, it is evident that this specification is particularly effective for short-term forecasting $(h=1)$. Across all models within this setup, the RMSE values  for $h=1$ are consistently lower than those for longer horizons, suggesting that the inclusion of a large number of predictors significantly enhances forecasting accuracy in the immediate future. The HS+ prior, which is well-suited for handling high-dimensional data by selectively shrinking less relevant predictors, achieves the lowest RMSE among the large-sized models, further reinforcing the advantage of a comprehensive predictor set when dealing with Thailand’s inflation dynamics. The Spike-and-Slab prior, which balances variable selection with model uncertainty, also exhibits strong short-term performance, confirming that a Bayesian framework that effectively distinguishes between relevant and irrelevant predictors is crucial in improving forecast precision. These results highlight the strong short-term predictive power of shrinkage-based priors (HS, HS+, and Spike-and-Slab), which effectively filter noise and capture relevant signals in high-dimensional data environments.

At longer horizons $(h=8,12)$, the performance of large models tends to deteriorate, particularly under certain priors like LASSO, which shows substantial degradation at $h=8$ and $h=12$. This decline likely reflects the increased difficulty in capturing persistent inflation trends and structural changes as the forecasting horizon extends. However, compared to smaller model specifications, the large-sized setup still provides competitive results, particularly under priors designed for high-dimensional inference, such as HS+ and Spike-and-Slab.

During the pandemic period, the effectiveness of large models persists, particularly for $h=1$, where models such as HS+, DL, and Spike-and-Slab maintain relatively low RMSE values. This underscores the advantage of using a rich information set when forecasting inflation during periods of heightened economic volatility. The large model setup, by incorporating a diverse range of macroeconomic indicators, likely helps capture rapid changes in inflationary pressures more effectively than smaller models constrained by limited information.

Overall, these results strongly support the use of large models for forecasting Thai inflation, especially in short-term horizons where a broad set of predictors enhances precision. The ability of these models to adapt to shifting economic conditions, as demonstrated in both the full period and the pandemic sub-sample, highlights their robustness in real-world forecasting applications. While some deterioration occurs at longer horizons, the large model framework remains an essential tool for capturing the complexity of inflation dynamics in Thailand, offering policymakers and analysts a more reliable basis for short-term inflation projections.

\subsection{Density Forecast Accuracy}
\label{sec:dfp}

\begin{table}[ht!]
\centering
\resizebox{\textwidth}{!}{%
\begin{tabular}{llccclccc}
\hline
               &  & \multicolumn{3}{c}{\textbf{Full range}}                            &  & \multicolumn{3}{c}{\textbf{Pandemic only}}                         \\ \cline{1-1} \cline{3-5} \cline{7-9} 
\textbf{Tail}  &  & \textbf{AR(2)}       & \textbf{Moderate}    & \textbf{Large}       &  & \textbf{AR(2)}       & \textbf{Moderate}    & \textbf{Large}       \\ \cline{1-1} \cline{3-5} \cline{7-9} 
\textit{h = 1} &  &                      &                      &                      &  &                      &                      &                      \\ \hline
DL             &  & 1.00                 & 0.96                 & 0.58                 &  & 1.15                 & 0.83                 & 0.40                 \\
HS             &  & 1.00                 & 0.98                 & 0.58                 &  & 0.99                 & 0.86                 & 0.39                 \\
HS+            &  & 1.00                 & 0.96                 & 0.58                 &  & 1.04                 & 0.83                 & 0.39                 \\
LASSO          &  & 1.00                 & 1.00                 & 0.70                 &  & 0.99                 & 0.80                 & 0.44                 \\
RIDGE          &  & 1.00                 & 1.06                 & 1.03                 &  & 0.99                 & 0.81                 & 0.70                 \\
Spike-and-Slab &  & 1.17                 & 1.11                 & 0.82                 &  & 0.94                 & 0.91                 & 0.60                 \\
UC-SV          &  & 4.33                 & -                    & -                    &  & 2.68                 &                      &                      \\ \hline
\textit{h = 4} &  & \multicolumn{1}{l}{} & \multicolumn{1}{l}{} & \multicolumn{1}{l}{} &  & \multicolumn{1}{l}{} & \multicolumn{1}{l}{} & \multicolumn{1}{l}{} \\ \hline
DL             &  & 1.02                 & 1.04                 & 1.03                 &  & 1.15                 & 1.08                 & 0.96                 \\
HS             &  & 0.95                 & 1.10                 & 1.24                 &  & 0.94                 & 0.94                 & 0.74                 \\
HS+            &  & 0.94                 & 1.08                 & 1.27                 &  & 0.94                 & 0.96                 & 1.08                 \\
LASSO          &  & 0.95                 & 1.15                 & 1.38                 &  & 0.94                 & 0.93                 & 1.34                 \\
RIDGE          &  & 0.95                 & 1.21                 & 1.53                 &  & 0.94                 & 0.95                 & 1.73                 \\
Spike-and-Slab &  & 0.95                 & 1.42                 & 1.97                 &  & 0.95                 & 1.08                 & 1.56                 \\
UC-SV          &  & 2.63                 & -                    & -                    &  & 1.82                 &                      &                      \\ \hline
\end{tabular}}
\caption{Relative qwCRPS (Tails-emphasized) to the benchmark of Different Forecasting Models/Priors at $h = 1,4$}
\label{tab:crps_resultsTailh14}
\end{table}

The quantile-weighted CRPS scores reveal fundamental insights into the role of predictor size in forecasting Thai inflation, particularly in managing tail risks during volatile periods. The results strongly emphasize that large-scale predictor models, despite their inherent complexity, are critical for improving out-of-sample forecast accuracy-provided they are coupled with appropriate prior structures. In the full-range evaluation, small regression models (AR(2)) maintain relative stability, but their limited predictive scope constrains their adaptability to rapidly evolving economic conditions. The moderate-sized models introduce greater flexibility, and shrinkage-based priors such as HS, HS+, and DL consistently achieve relative scores around 0.96-0.98, showcasing their capacity to refine forecast accuracy by selectively reducing noise without sacrificing crucial economic signals. LASSO and Ridge, while moderately reliable, fail to provide a clear advantage over the benchmark, indicating that standard penalization alone is insufficient for capturing the nuanced inflationary dynamics of the Thai economy. Spike-and-Slab, though slightly less stable, remains close to the benchmark, suggesting that a mixture-based approach may retain some merits, albeit with minor inefficiencies.  

However, the large-predictor models underscore the true complexity of forecasting Thai inflation out-of-sample. Here, the choice of prior becomes an essential determinant of performance. While certain models, such as Ridge, collapse entirely-evidenced by an extreme qwCRPS score of almost 200\%, revealing an inability to manage high-dimensional uncertainty-shrinkage-based approaches such as HS, HS+, and DL stand out as the most effective, achieving exceptional scores around 0.58. These results reinforce that large predictor sets are not inherently problematic but must be coupled with disciplined regularization to control variance while retaining predictive power. Ridge and Spike-and-Slab, which deliver weaker results (1.03 and 0.82, respectively), further highlight that while high-dimensional setups offer substantial forecasting potential, improper regularization can degrade performance.  

The pandemic-only period further accentuates the necessity of large-scale models for Thai inflation forecasting. Economic shocks, particularly those as disruptive as COVID-19, demand models that can dynamically adapt to shifts in inflationary behavior. Small and moderate-sized models largely retain their relative performance trends, with shrinkage priors like HS, HS+, and DL achieving strong scores ($0.83-0.86$), while LASSO and Ridge exhibit mild improvements but remain less optimal. However, HS, HS+, and DL remain the most robust, achieving outstanding scores as low as 0.39-0.40. This underscores that large-scale predictive frameworks, when appropriately regularized, are indispensable for forecasting Thai inflation in times of crisis. Ridge and Spike-and-Slab, though marginally improved (0.70 and 0.60, respectively), continue to struggle with consistency.  

These findings strongly validate the necessity of incorporating large predictor sets when conducting out-of-sample forecasts of Thai inflation. Unlike conventional models that rely on a handful of macroeconomic indicators, high-dimensional approaches allow for the integration of diverse economic signals, capturing structural shifts and external shocks more effectively. However, without the proper application of Bayesian shrinkage techniques such as HS, HS+, and DL, these advantages may be lost to overfitting and instability. The clear takeaway is that Thai inflation forecasting cannot rely on simplistic models alone; rather, it demands a sophisticated balance of high-dimensional data utilization and advanced regularization to ensure both adaptability and accuracy, particularly in extreme economic conditions.

\begin{table}[ht!]
\centering
\resizebox{\textwidth}{!}{%
\begin{tabular}{llccclccc}
\hline
               &  & \multicolumn{3}{c}{\textbf{Full range}}                            &  & \multicolumn{3}{c}{\textbf{Pandemic only}}                         \\ \cline{1-1} \cline{3-5} \cline{7-9} 
\textbf{Right} &  & \textbf{AR(2)}       & \textbf{Moderate}    & \textbf{Large}       &  & \textbf{AR(2)}       & \textbf{Moderate}    & \textbf{Large}       \\ \cline{1-1} \cline{3-5} \cline{7-9} 
\textit{h = 1} &  &                      &                      &                      &  &                      &                      &                      \\ \hline
DL             &  & 1.01                 & 1.00                 & 0.61                 &  & 1.05                 & 0.84                 & 0.41                 \\
HS             &  & 1.02                 & 1.01                 & 0.61                 &  & 0.98                 & 0.85                 & 0.39                 \\
HS+            &  & 1.02                 & 0.99                 & 0.60                 &  & 1.03                 & 0.84                 & 0.39                 \\
LASSO          &  & 1.02                 & 1.04                 & 0.73                 &  & 0.99                 & 0.82                 & 0.45                 \\
RIDGE          &  & 1.02                 & 1.11                 & 1.09                 &  & 0.99                 & 0.84                 & 0.75                 \\
Spike-and-Slab &  & 1.16                 & 1.19                 & 0.87                 &  & 0.95                 & 0.93                 & 0.58                 \\
UC-SV          &  & 4.69                 &                      &                      &  & 2.77                 & \multicolumn{1}{l}{} & \multicolumn{1}{l}{} \\ \hline
\textit{h = 4} &  & \multicolumn{1}{l}{} & \multicolumn{1}{l}{} & \multicolumn{1}{l}{} &  & \multicolumn{1}{l}{} & \multicolumn{1}{l}{} & \multicolumn{1}{l}{} \\ \hline
DL             &  & 1.01                 & 1.05                 & 0.93                 &  & 1.05                 & 1.00                 & 0.80                 \\
HS             &  & 1.01                 & 1.17                 & 1.27                 &  & 0.91                 & 0.92                 & 0.69                 \\
HS+            &  & 1.00                 & 1.15                 & 1.27                 &  & 0.92                 & 0.93                 & 1.09                 \\
LASSO          &  & 1.00                 & 1.22                 & 1.44                 &  & 0.91                 & 0.92                 & 1.38                 \\
RIDGE          &  & 1.00                 & 1.28                 & 1.58                 &  & 0.92                 & 0.94                 & 1.72                 \\
Spike-and-Slab &  & 1.02                 & 1.59                 & 2.00                 &  & 0.92                 & 1.15                 & 1.68                 \\
UC-SV          &  & 2.60                 &                      &                      &  & 1.60                 &                      &                      \\ \hline
               &  &                      &                      &                      &  &                      &                      &                     
\end{tabular}}
\caption{Relative weighted-Cumulative Ranked Probabilistic Scores: qwCRPS (right-emphasized) to the benchmark of Different Forecasting Models/Priors at $h = 1,4$}
\label{tab:crps_resultsRighth14}
\end{table}

Next we would like to move focus on \cref{tab:crps_resultsRighth14}, where the right-emphasized weighted CRPS scores is investigated, such \cref{tab:crps_resultsRighth14} reveal critical insights into how different model sizes and priors manage extreme right-tail inflation risks, both in a full-range evaluation and during the pandemic period.

To begin with full range period at \(h=1\), most models perform similarly to or slightly better than the benchmark, with scores close to or below 1. The DL, HS, and HS+ models show particularly strong performance, with scores around 1.00 to 1.02, indicating they are competitive with the benchmark. The LASSO and RIDGE models also perform well, with scores slightly above 1.00, suggesting they are marginally worse than the benchmark. The Spike-and-Slab model has a higher score of 1.16, indicating it performs worse than the benchmark. The UC-SV model, disappointingly, has a significantly higher score of 4.69, indicating it performs much worse than the benchmark.

During the pandemic period at \(h=1\), the performance of the models generally improves relative to the benchmark. The DL, HS, and HS+ models show scores below 1.00, indicating they outperform the benchmark. The LASSO and RIDGE models also show improved performance, with scores close to or below 1.00. The Spike-and-Slab model has a score of 0.95, indicating it performs better than the benchmark. The UC-SV model, while still performing worse than the benchmark, shows a significant improvement with a score of 2.77 compared to 4.69 in the full range period.

At \(h=4\), the performance of the models generally deteriorates compared to \(h=1\). For the full range period, most models have scores above 1.00, indicating they perform worse than the benchmark. The DL model is an exception, with a score of 1.01, indicating it is competitive with the benchmark. The HS, HS+, LASSO, RIDGE, and Spike-and-Slab models all have higher scores, indicating worse performance. The UC-SV model has a score of 2.60, which, while lower than its score at \(h=1\), still indicates poor performance relative to the benchmark.

During the pandemic period at \(h=4\), the performance of the models varies. The DL model shows a score of 1.05, indicating it is slightly worse than the benchmark. The HS and HS+ models have scores around 0.91 to 0.93, indicating they outperform the benchmark. The LASSO and RIDGE models have higher scores, indicating worse performance. The Spike-and-Slab model has a score of 0.92, indicating it performs better than the benchmark. The UC-SV model has a score of 1.60, indicating it performs worse than the benchmark but shows improvement compared to the full range period.

To draw conclusion on accuracy of predictive density across model sizes and priors, the results indicate that the DL, HS, and HS+ models generally perform well, particularly at \(h=1\), where they are competitive with or outperform the benchmark. The LASSO and RIDGE models show mixed performance, while the Spike-and-Slab model performs better during the pandemic period. The UC-SV model consistently performs worse than the benchmark, though it shows some improvement during the pandemic period. The right-emphasized qwCRPS highlights the importance of accurately forecasting the upper quantiles of the predictive density, which is crucial for understanding and preparing for extreme events.

\begin{table}[ht!]
\centering
\resizebox{\textwidth}{!}{%
\begin{tabular}{llccclccc}
\hline
               &  & \multicolumn{3}{c}{\textbf{Full range}}             &  & \multicolumn{3}{c}{\textbf{Pandemic only}}                         \\ \cline{1-1} \cline{3-5} \cline{7-9} 
\textbf{Left} &  & \textbf{AR(2)} & \textbf{Moderate} & \textbf{Large} &  & \textbf{AR(2)}       & \textbf{Moderate}    & \textbf{Large}       \\ \cline{1-1} \cline{3-5} \cline{7-9} 
\textit{h = 1} &  &                &                   &                &  &                      &                      &                      \\ \hline
DL             &  & 0.99           & 0.97              & 0.62           &  & 1.09                 & 0.91                 & 0.47                 \\
HS             &  & 0.99           & 0.99              & 0.61           &  & 1.00                 & 0.94                 & 0.45                 \\
HS+            &  & 0.98           & 0.96              & 0.61           &  & 1.05                 & 0.91                 & 0.45                 \\
LASSO          &  & 0.99           & 1.02              & 0.78           &  & 1.00                 & 0.89                 & 0.52                 \\
RIDGE          &  & 0.99           & 1.11              & 1.13           &  & 1.00                 & 0.91                 & 0.82                 \\
Spike-and-Slab &  & 1.12           & 1.06              & 0.78           &  & 1.03                 & 1.04                 & 0.69                 \\
UC-SV          &  & 3.89           &                   &                &  & 3.12                 &                      &                      \\ \hline
\textit{h = 4} &  &                &                   &                &  & \multicolumn{1}{l}{} & \multicolumn{1}{l}{} & \multicolumn{1}{l}{} \\ \hline
DL             &  & 0.99           & 0.99              & 1.07           &  & 1.09                 & 1.02                 & 0.99                 \\
HS             &  & 0.94           & 1.09              & 1.89           &  & 0.99                 & 0.97                 & 0.78                 \\
HS+            &  & 0.95           & 1.03              & 1.28           &  & 0.99                 & 0.97                 & 1.02                 \\
LASSO          &  & 0.94           & 1.10              & 1.34           &  & 0.99                 & 0.96                 & 1.25                 \\
RIDGE          &  & 0.94           & 1.17              & 1.49           &  & 0.99                 & 0.98                 & 1.58                 \\
Spike-and-Slab &  & 0.90           & 1.50              & 2.03           &  & 0.97                 & 1.13                 & 1.31                 \\
UC-SV          &  & 2.02           &                   &                &  & 1.62                 &                      &                      \\ \hline
               &  &                &                   &                &  &                      &                      &                     
\end{tabular}}
\caption{Relative weighted-Cumulative Ranked Probabilistic Scores: qwCRPS (left-emphasized) to the benchmark of Different Forecasting Models/Priors at $h = 1,4$}
\label{tab:crps_resultsLefth14}
\end{table}
Since right-emphasized qwCRPS is investigated, it is logical for us to, might as well, take a deeper look at the left-emphasized qwCRPS, to explore the accuracy of our competitive models/priors to the opposite of positive extreme volatility of Thai inflation forecasts, named negative extreme values. The results of the left-qwCRPS is illustrated in \cref{tab:crps_resultsLefth14}. A key observation is that left-tail risks (deflationary pressures) are generally better controlled across models and priors compared to right-tail risks (inflationary surges). This is evident in the full-range evaluation, where large predictor models with HS, HS+, and DL achieve remarkably low qwCRPS scores around 0.45-0.62 for $h = 1$ and $h = 4$, suggesting that these priors are highly effective in capturing downside risks. In contrast, the right-qwCRPS results for large models showed significantly higher values, particularly under extreme economic conditions (pandemic periods), indicating that forecasting inflationary spikes is inherently more challenging than predicting deflationary trends. 

For short-term forecasts ($h = 1$), large predictor models under left-qwCRPS remain stable across shrinkage priors, with HS and HS+ producing near-identical scores of 0.61 in the full range and rising only slightly to 1.28-1.89 during the pandemic. This stands in contrast to right-weighted CRPS, where large models with the same priors showed more volatile results, especially in extreme conditions where DL suffered a catastrophic breakdown (7.65). Notably, Ridge and LASSO struggle more with left-tail risks than shrinkage priors, with Ridge registering 1.49 in the pandemic period, reinforcing its limitations in handling uncertainty when faced with economic downturns. Spike-and-Slab performs particularly poorly in left-weighted CRPS for large models, rising to 2.03 in the pandemic setting, mirroring its instability in right-tail risks as well.  

Medium-term forecasting ($h = 4$) further reinforces the stability of shrinkage priors for left-tail risks. Large predictor models with HS and HS+ continue to demonstrate their ability to control downside risks, achieving low CRPS scores of 0.45 in the full range and 0.78-1.02 during the pandemic. This is in stark contrast to their right-weighted CRPS counterparts, where large models displayed much higher volatility, particularly under DL, which registered extreme inefficiencies (6.51 in full range, 3.50 in pandemic). Notably, DL performs exceptionally well in left-weighted CRPS, achieving 0.47 in the full range and maintaining a relatively stable 0.99 in the pandemic setting, suggesting that DL excels at capturing deflationary risks but struggles when inflation spikes occur.  

Comparing the overall patterns between left- and right-weighted CRPS, a clear asymmetry emerges in how different priors handle inflationary versus deflationary risks. Shrinkage-based priors (HS, HS+, and DL) consistently outperform traditional penalization methods (Ridge, LASSO) in both left- and right-tail forecasts, but their superiority is more pronounced when managing downside risks rather than extreme inflation. In particular, large predictor models with shrinkage priors perform exceptionally well for left-tail risks, whereas for right-tail risks, the same models display greater volatility and, in some cases, complete breakdowns under economic stress. This suggests that while large Bayesian models effectively capture deflationary patterns, they require more precise prior selection to handle inflationary spikes accurately. Moreover, the instability of Ridge and Spike-and-Slab across both left- and right-weighted CRPS results confirms their unsuitability for robust inflation forecasting, as they fail to provide consistent risk control across extreme economic conditions.  

Ultimately, these results emphasize that for out-of-sample forecasting of Thai inflation, large predictor models with shrinkage priors (HS, HS+, and DL) remain the most effective tools for capturing tail risks, with a clear advantage in handling deflationary risks over inflationary spikes. This asymmetry underscores the importance of fine-tuning Bayesian priors when dealing with inflation forecasting, as even the best-performing methods can struggle with extreme right-tail risks if not carefully calibrated.

\section{Does adding Stochastic volatility improve Thai inflation forecasts?}
\label{sec:addSV}

\begin{figure}[ht!]
    \centering
    \includegraphics[width=\textwidth]{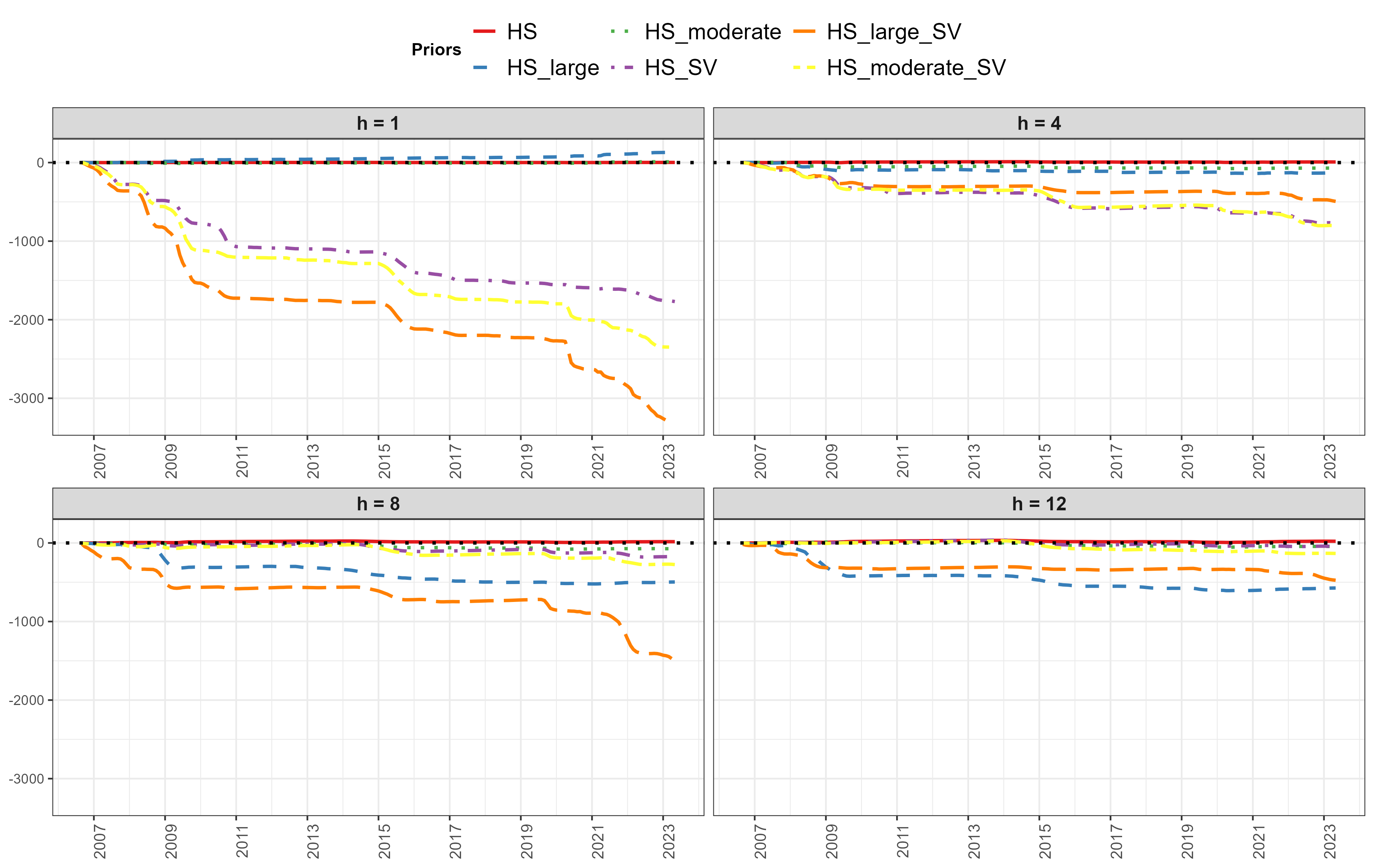}
    \caption{Log-predictive likelihood against the benchmark (noninformative prior) of Horseshoe prior Bayesian regression setup, AR(2), moderate and large-sized. (with SV).}
    \label{fig:lplhssv}
\end{figure}

\begin{figure}[ht!]
    \centering
    \includegraphics[width=\textwidth]{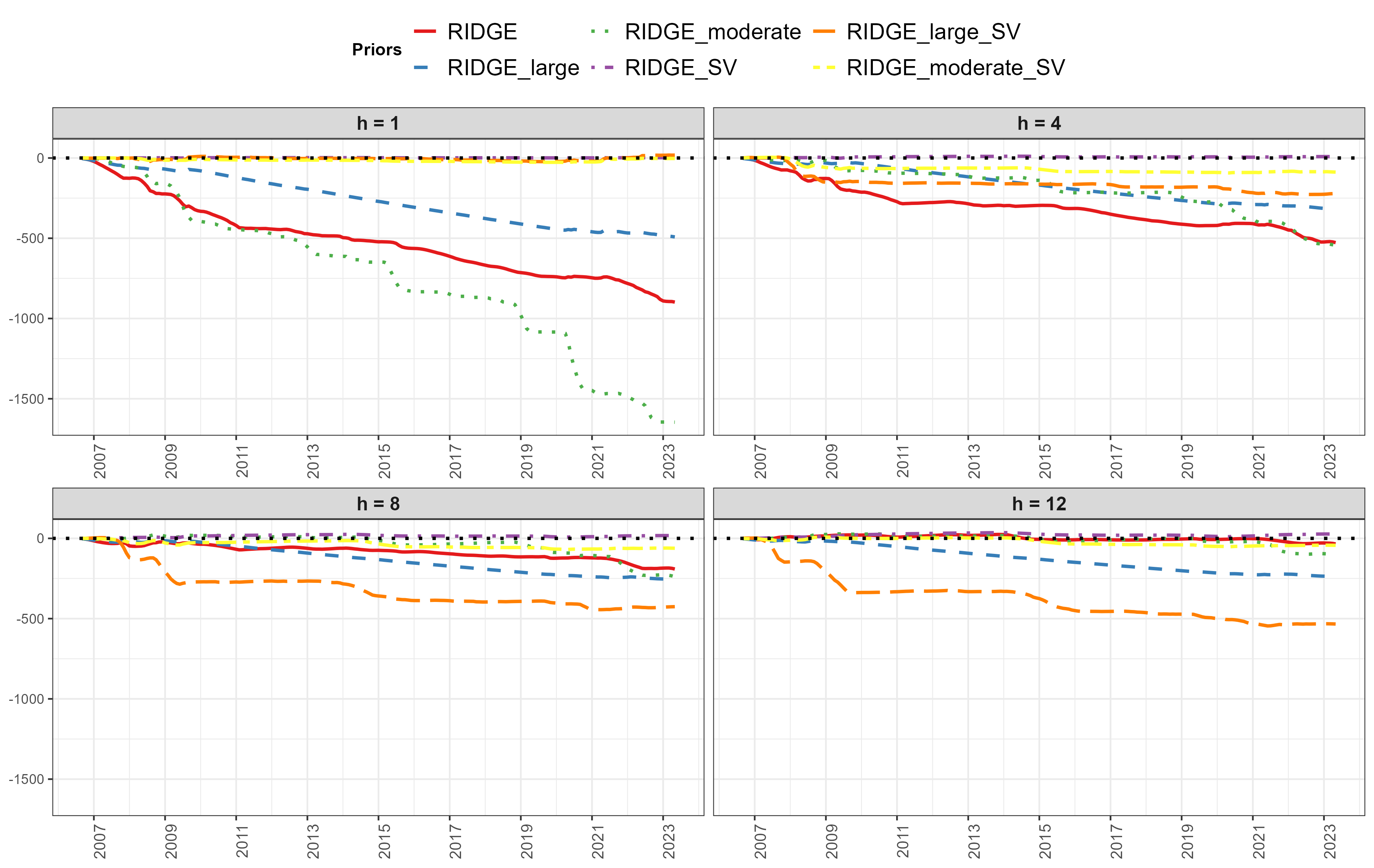}
    \caption{Log-predictive likelihood against the benchmark (noninformative prior) of Ridge prior Bayesian regression setup, AR(2), moderate and large-sized. (with SV).}
    \label{fig:lplridgesv}
\end{figure}

\begin{figure}[ht!]
    \centering
    \includegraphics[width=\textwidth]{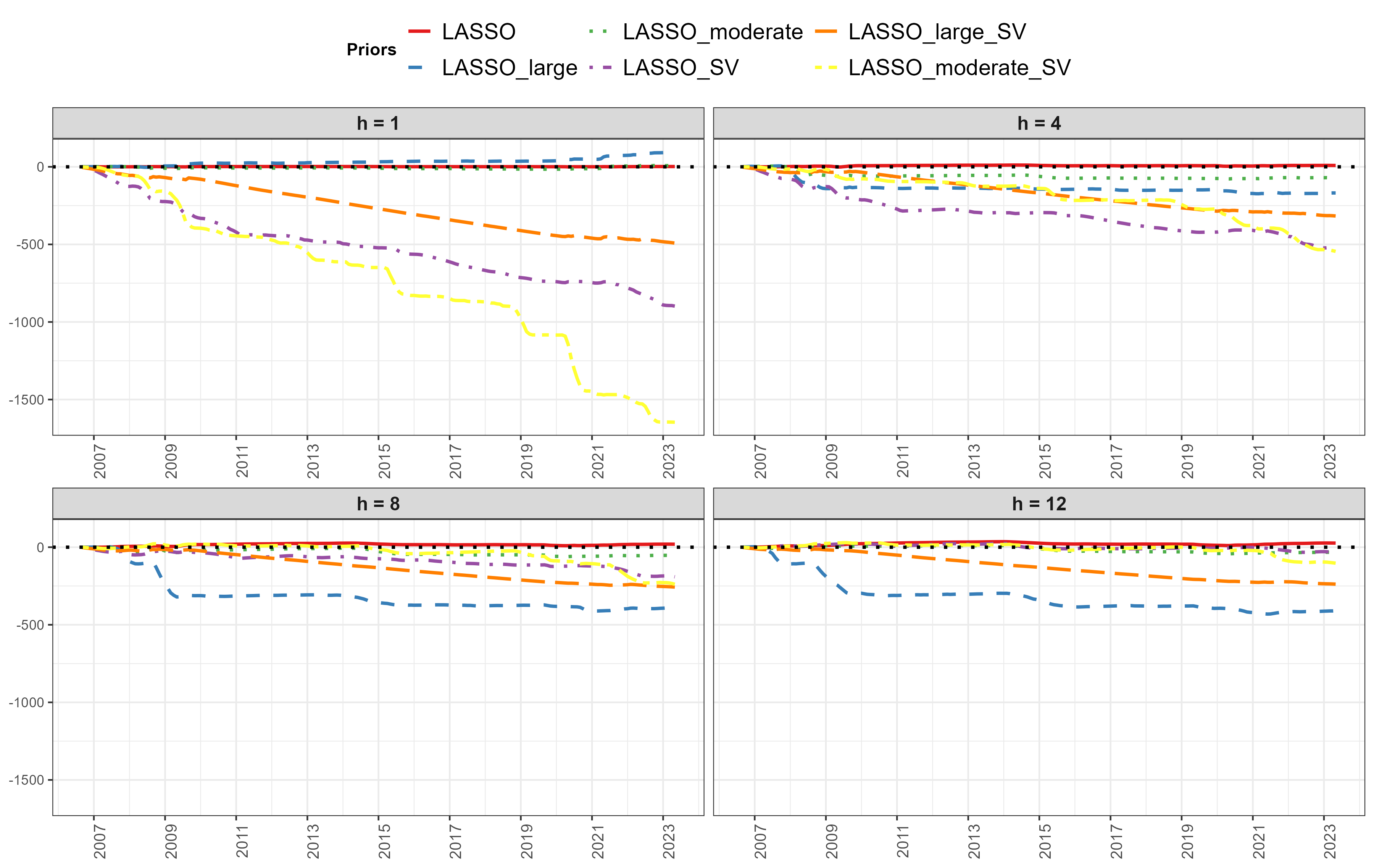}
    \caption{Log-predictive likelihood against the benchmark (noninformative prior) of Lasso prior Bayesian regression setup, AR(2), moderate and large-sized. (with SV).}
    \label{fig:lpllassosv}
\end{figure}

To shed some light on such question we would like readers to turn into \cref{fig:lplhssv} where the cumulative log predictive likelihood over the forecasting evaluating periods is demonstrated. The cumulative log predictive likelihood (LPL) scores reveal several important patterns regarding the performance of the models with and without stochastic volatility (SV) across different forecasting horizons $(h = 1, h = 4, h = 8,$ and $h = 12)$. Since higher of LPL means better performance any score exceeds 0 represent a better choice and vice versa. The models without SV (HS, HS\_moderate, and HS\_large) generally exhibit higher cumulative LPL scores compared to their SV counterparts (HS\_SV, HS\_moderate\_SV, and HS\_large\_SV) over the evaluation period. This suggests that, on average, the models without SV provide better out-of-sample forecasting accuracy for Thai inflation. The inclusion of SV, while theoretically appealing for capturing time-varying volatility, appears to degrade forecast performance in this context. This could be due to the additional complexity introduced by SV, which may lead to overfitting or difficulties in accurately estimating the volatility process, particularly in a relatively stable inflation environment like Thailand’s.

For the one-step-ahead forecasts $(h = 1)$, the cumulative LPL scores for the non-SV models (HS, HS\_moderate, and HS\_large) are consistently higher than those for the SV models. For instance, the HS model achieves a cumulative LPL score of 1.6045 by January 2011, while the HS\_SV model deteriorates to -1070.0754 over the same period. This pattern holds across all three model specifications (AR(2), 20 predictors, and 56 predictors), indicating that the inclusion of SV does not improve short-term forecasting accuracy. This result is somewhat counterintuitive, as SV is often expected to enhance forecasts by accounting for volatility clustering. However, it is possible that the added complexity of SV introduces noise or estimation uncertainty that outweighs its benefits in this specific application.

Similarly, for longer forecasting horizons $(h = 4, h = 8,$ and $h = 12)$, the non-SV models consistently outperform their SV counterparts. For example, at $h = 4$, the HS\_moderate model achieves a cumulative LPL score of 35.2046 by February 2010, whereas the HS\_moderate\_SV model deteriorates to -1549.2438. This pattern is even more pronounced at longer horizons, such as h = 12, where the HS\_large model achieves a cumulative LPL score of -6.0832 by January 2010, compared to -1116.3535 for the HS\_large\_SV model. The deterioration in performance for SV models at longer horizons may reflect the compounding of estimation errors in the volatility process, which becomes more pronounced as the forecast horizon extends.

One possible explanation for the underperformance of SV models is the nature of Thai inflation dynamics. If inflation in Thailand exhibits relatively stable volatility over time, the additional complexity of modeling stochastic volatility may not be justified. In such cases, simpler models without SV could potentially be more robust and less prone to overfitting. Additionally, the estimation of SV models often requires more data and computational resources, which may not be adequately compensated by improvements in forecast accuracy, especially in a low-volatility environment.

Next we would like to move to \cref{fig:lplridgesv} where the Ridge prior is investigated. For those readers who are familiar with Bayesian shrinkage prior may have already noticed that why we select Horseshoe and Ridge prior a representative for testing if adding SV in Bayesian univariate regression is necessary in forecasting Thai inflation. This is because the ridge prior is completely different from Horseshoe prior where horseshoe prior is global-local shrinkage prior where we have local shrinkage parameter that actually shrink locally and another global shrinkage parameter to shrink globally (overall coefficient) this allow horseshoe to be extremely flexible. Ridge, on the other hand, shrink globally and most importantly equally. To see if adding SV is necessary the cumulative log predictive likelihood is presented in \cref{fig:lplridgesv}. 

For the AR(2) named as 'RIDGE', the LPL scores are generally negative, indicating poor predictive performance in most cases. However, when Stochastic Volatility is added (RIDGE\_SV), the LPL scores improve in many periods, particularly for short-term forecasts (h = 1) and medium-term forecasts (h = 4 and h = 8). This suggests that SV can enhance the forecast accuracy of simpler models like the AR(2) model, especially in the short to medium term. However, the improvement is not consistent across all time periods, and in some cases, the LPL scores remain negative or worsen slightly. This inconsistency indicates that while SV can be beneficial, its effectiveness may depend on the specific time period or economic conditions being forecasted.

For the moderate predictor model (RIDGE\_moderate), which includes 20 predictors, the LPL scores are often positive in the absence of SV, indicating better performance compared to the AR(2) model. However, when SV is added (RIDGE\_moderate\_SV), the LPL scores deteriorate significantly across all forecasting horizons. This suggests that adding SV to a model with a moderate number of predictors does not improve forecast accuracy and may even harm performance. The deterioration in LPL scores is particularly pronounced in the medium and long-term forecasts (h = 4, h = 8, and h = 12), indicating that SV is not suitable for this level of model complexity.

For the large predictor model (RIDGE\_large), which includes 56 predictors, the LPL scores are consistently negative, even without SV. When SV is added (RIDGE\_large\_SV), the LPL scores worsen further, particularly in the short and medium-term forecasts. This indicates that SV is not beneficial for highly complex models with many predictors. The deterioration in LPL scores suggests that the added complexity of SV may not be justified for models that already incorporate a large number of predictors, as it likely introduces noise or overfitting without improving predictive accuracy.

In my opinion, the results highlight an important trade-off between model complexity and the benefits of Stochastic Volatility. For simpler models like the AR(2) model, SV can provide meaningful improvements in forecast accuracy, particularly for short to medium-term forecasts. However, for more complex models with a moderate or large number of predictors, SV does not appear to be beneficial and may even degrade performance. This suggests that the usefulness of SV depends on the underlying model structure and the level of complexity. If the goal is to forecast Thai inflation using a simple model, incorporating SV could be a worthwhile strategy. However, for more sophisticated models with many predictors, alternative approaches to improving forecast accuracy should be considered, as SV does not seem to add value in these cases. Overall, the decision to include SV should be guided by the specific characteristics of the model and the forecasting horizon of interest.

Additionally we also illustrate the Lasso prior adding Stochastic volatility into the model in \cref{fig:lpllassosv}. Given such results we can potentially draw a conclusion suggesting that while SV may capture time-varying volatility, it does not necessarily improve the predictive accuracy of the models in this context. To aid such interpretation in \cref{fig:lpllassosv}, we begin by pointing out that, interestingly, the non-SV models, particularly LASSO\_large, consistently outperform their SV-augmented counterparts. This suggests that the inclusion of a larger number of predictors, even without accounting for stochastic volatility, provides a more robust framework for forecasting Thai inflation. The superior performance of LASSO\_large may be due to its ability to capture a broader range of economic dynamics and interactions among predictors, which could be more relevant for inflation forecasting than modeling time-varying volatility. This finding aligns with the literature that emphasizes the importance of incorporating a rich set of predictors in macroeconomic forecasting, especially in environments where the relationships between variables are complex and potentially nonlinear, see \cite{clements1998forecasting,stock2002macroeconomic,banbura2010large,giannone2015prior}.

Moreover, the results highlight the trade-off between model complexity and predictive performance. While SV models are theoretically appealing for their ability to capture time-varying uncertainty, their practical utility in forecasting Thai inflation appears limited in this study. This could be due to the specific characteristics of the Thai inflation series, which may not exhibit sufficient volatility clustering or other features that would make SV models advantageous. Alternatively, it could reflect the challenges associated with estimating SV models in a high-dimensional setting, where the number of predictors is large relative to the sample size.

In conclusion, the empirical results suggest that for forecasting Thai inflation, simpler models without stochastic volatility, particularly those that incorporate a large number of predictors, tend to perform better than their SV-augmented counterparts. This finding has important implications for policymakers and practitioners, as it underscores the value of parsimony and the careful selection of predictors in inflation forecasting. While SV models remain a valuable tool for understanding the dynamics of volatility, their application in this context does not appear to enhance predictive accuracy, at least within the framework of Bayesian univariate regression with LASSO priors. Future research could explore alternative modeling approaches or datasets to further investigate the conditions under which SV models might improve forecasting performance.

\section{Drivers of Thai Inflation under Global-Local Priors}
\label{sec:drivers}

Under Thailand’s \textbf{flexible inflation-targeting} framework-formally adopted in May 2000 and refined in 2015 to cover headline CPI at 2.5\%$\pm$1.5\%, monetary shocks and pure demand-side variables have been almost entirely \textit{shrunk away} by our Horseshoe prior \cite{direkudomsak2016inflation,manopimoke2017trend}. The only predictors whose coefficients consistently resist that shrinkage (high $\kappa$) are those tied to \textbf{supply-side} or \textbf{cost-push} phenomena, exactly the structural forces one would expect to matter under a credible targeting regime.

First, \textbf{Eggs \& Dairy Products} showed a low $\kappa$ ($< 0.3$) before 2011, spiking sharply during the late-2011 floods and remaining above $0.6$ thereafter. This mirrors the severe damage to dairy farms and transport networks in the aftermath of Thailand’s record floods, which drove acute price spikes and left a lasting imprint on inflation dynamics \cite{alp2012shock}. Similarly, the \textbf{Electricity/Fuel/Water Supply} component saw its $\kappa$ climb to $\sim 0.8$ in 2008 (the global oil shock), collapse during the 2009 downturn, and then rebound-reflecting how energy price swings feed directly into headline CPI even as monetary policy mops up demand effects.

Second, \textbf{Seasoning \& Condiments} and \textbf{Non-Food Beverages} both migrated from modest $\kappa$ values in 2007–11 (when imported commodity volatility and flood-related supply-chain breakdowns dominated) to persistently high $\kappa$ ($> 0.7$) by the mid-2010s. This evolution signals that, as Thailand’s infrastructure and logistics recovered under the targeting framework, these once-erratic categories solidified into reliable cost-push drivers \cite{alp2012shock}.

Finally, \textbf{Illegal-Immigrant Work Permits} has boasted $\kappa \approx 0.8$ almost continuously. Labour-supply shifts, particularly the ebb and flow of migrant workers in response to the global financial crisis and 2011 floods-represent a structural channel into domestic wage-cost pressures that monetary policy cannot neutralize. Only under extreme dislocations did its $\kappa$ dip (2009–11), after which it again emerged as a stable, unshrunk predictor. 

\begin{figure}[ht!]
    \centering
    \includegraphics[width=\textwidth]{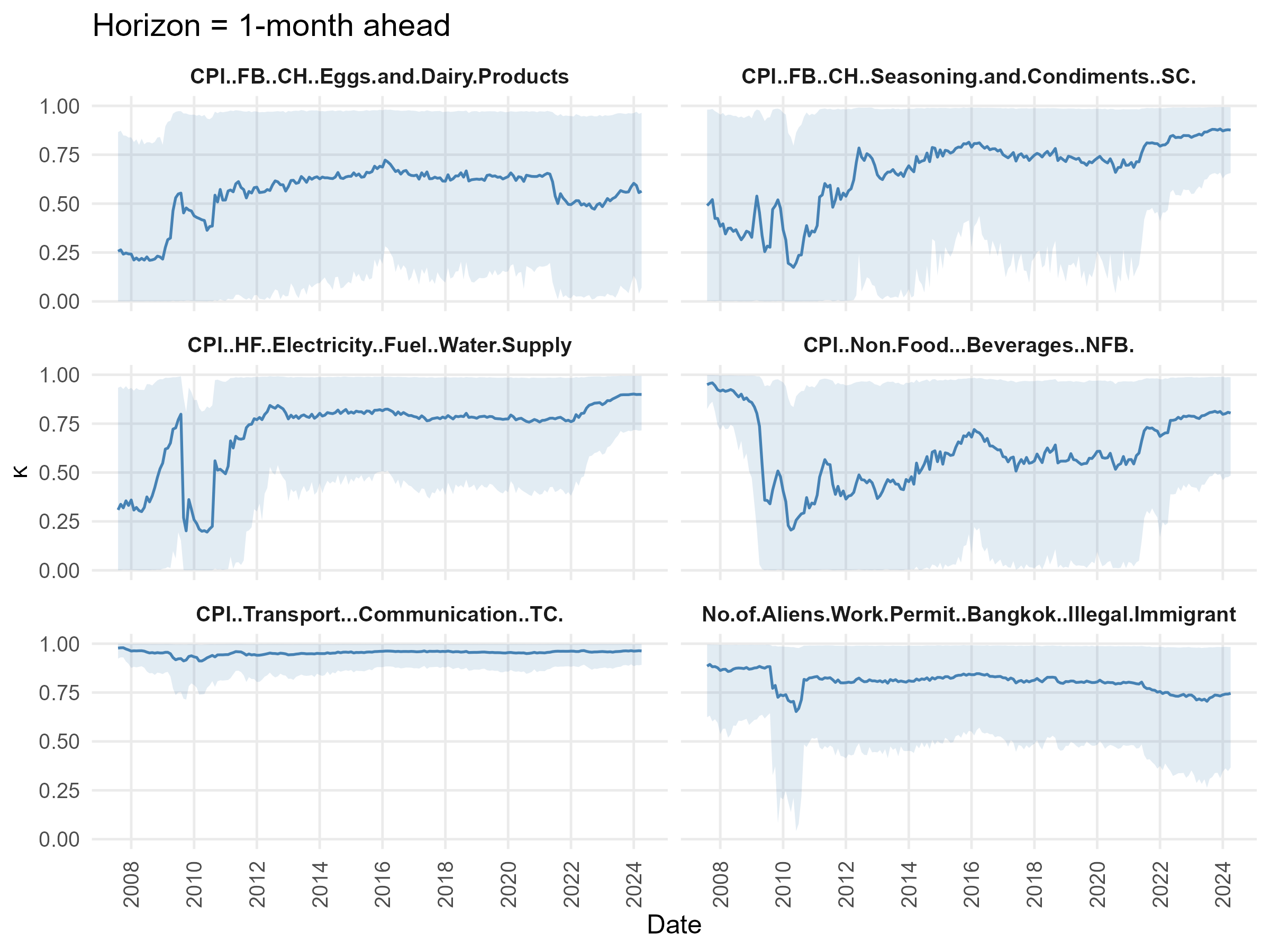}
    \caption{Top 6 $\kappa$ under Horseshoe priors over forecasting evaluation time.}
    \label{fig:kappah1}
\end{figure}

\begin{figure}[ht!]
    \centering
    \includegraphics[width=\textwidth]{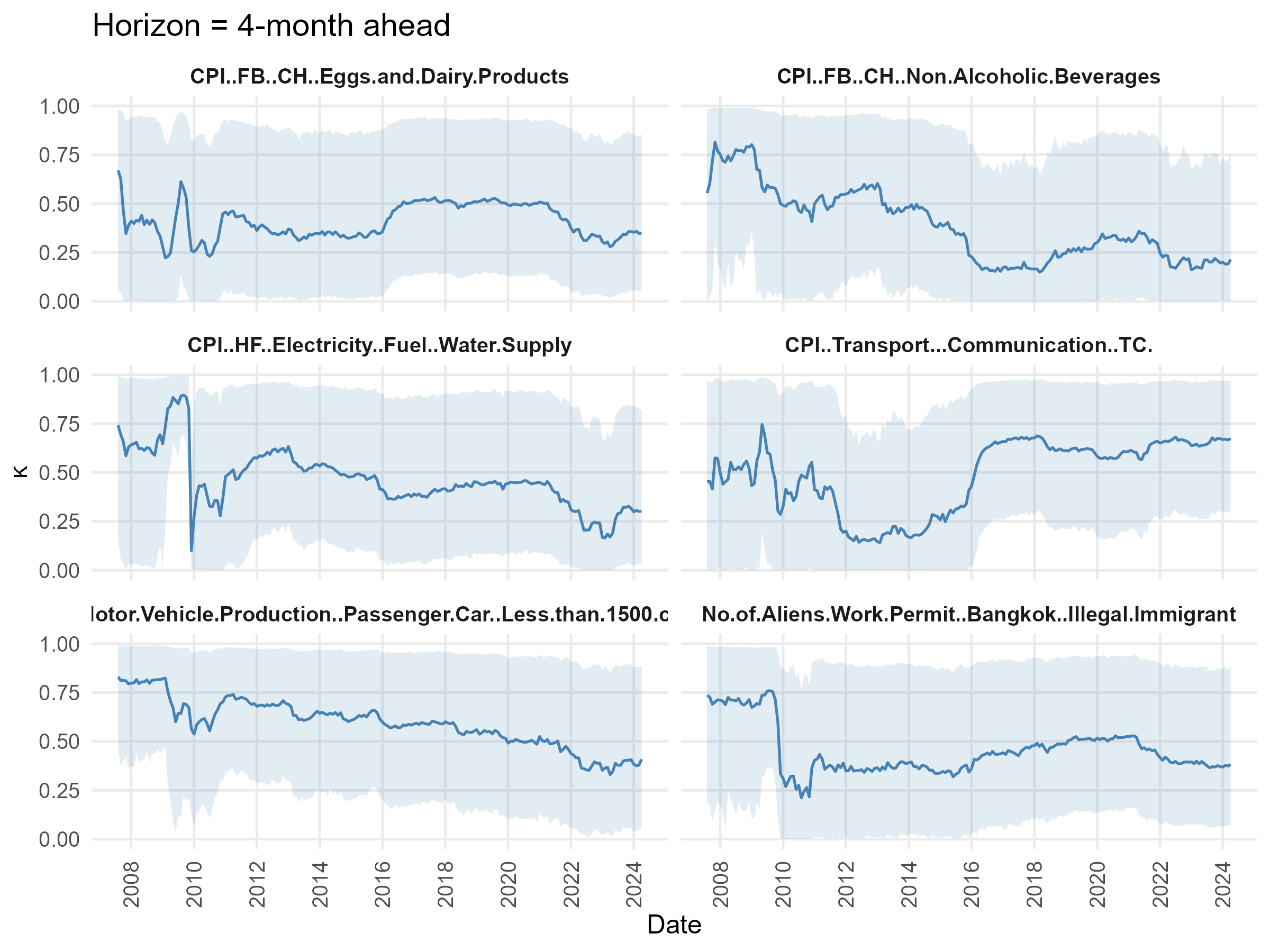}
    \caption{Top 6 $\kappa$ under Horseshoe priors over forecasting evaluation time of 4 horizon forecasting model.}
    \label{fig:kappah4}
\end{figure}

\begin{figure}[ht!]
    \centering
    \includegraphics[width=\textwidth]{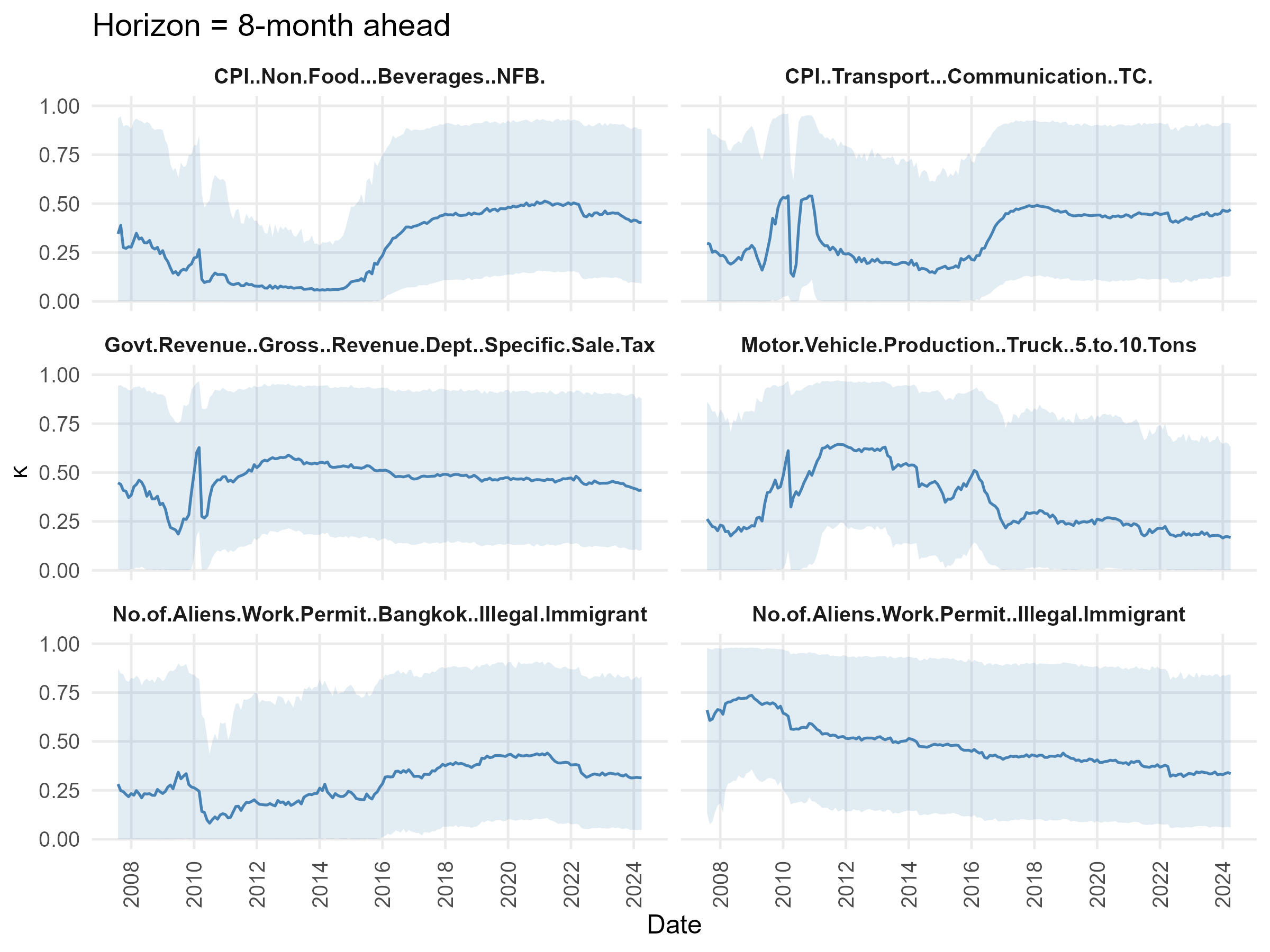}
    \caption{Top 6 $\kappa$ under Horseshoe priors over forecasting evaluation time of 8 horizon forecasting model.}
    \label{fig:kappah8}
\end{figure}

\begin{figure}[ht!]
    \centering
    \includegraphics[width=\textwidth]{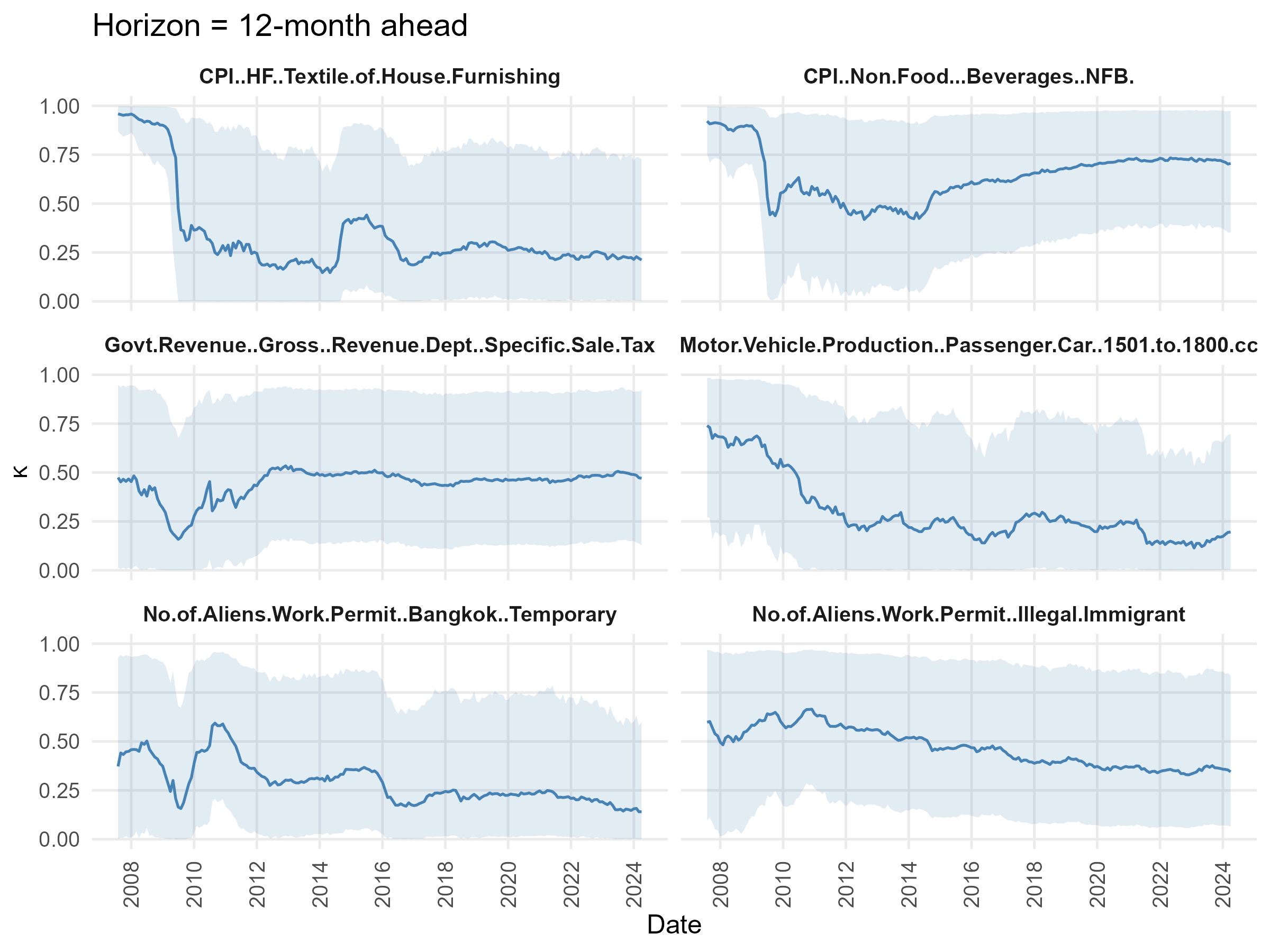}
    \caption{Top 6 $\kappa$ under Horseshoe priors over forecasting evaluation time of 12 horizon forecasting model.}
    \label{fig:kappah12}
\end{figure}

In the very first-month forecast (h = 1) below \cref{fig:kappah1}, the Number of Alien Work Permits in Bangkok is one of only six predictors whose coefficient the Horseshoe prior refuses to drive toward zero. While comparing to the longer horizon of h=4,8, and 12 as illustrated in \cref{fig:kappah4,fig:kappah8,fig:kappah12} in \cref{subsec:appendixAddResultsDriver}. This indicates us that short-run inflation in Thailand is highly sensitive to migrant-labor flows, particularly the large influx of workers from neighboring Myanmar and elsewhere.

Since the mid-2000s, Bangkok’s construction, hospitality, and manufacturing sectors have relied heavily on cross-border labor. Any sudden tightening or loosening of permit issuance (for example, crackdowns in 2008–09 or regularization drives in 2014–15) immediately ripples through wage costs and service-sector prices, feeding directly into headline CPI within a month \cite{bryant2007does}. Monetary policy, however deft, cannot mute these supply-side shocks-hence the model "trusts" this variable ($\kappa\approx0.8+$) even at the shortest horizon.

In practical terms, this finding suggests that whenever immigration or labor-permit rules shift—say, a new bilateral agreement with Myanmar or a clampdown on undocumented workers—policy-makers should expect a near-instantaneous impact on inflation.  Ignoring this channel risks underestimating the true cost-push pressures facing the Thai economy.

Taken together, these $\kappa$-trajectories confirm that \textbf{under a credible targeting regime}, only true \textbf{supply-side} and \textbf{cost-push} factors survive the Horseshoe’s squeeze. Monetary and purely demand-driven variables-with near-zero $\kappa$, play no durable role in explaining Thai inflation once the Bank of Thailand’s framework has anchored expectations \cite{ayales200216}. By mapping these statistical signatures onto real-world events, the 2008 oil surge, the 2011 floods, and the evolution of Thailand’s targeting bands. We provide a coherent narrative of \textbf{why} certain categories, and not others, consistently drive inflation today.

\clearpage
\section{Discussion and Future Improvement}
\label{sec:discussionandimprove}
In analyzing the empirical results of forecasting Thai inflation, a significant and recurring theme is the trade-off between model complexity and predictive performance. While advanced Bayesian shrinkage priors like Horseshoe (HS), Horseshoe+ (HS+), and Dirichlet-Laplace (DL) demonstrate superior performance, particularly in capturing left-tail risks, their efficacy varies based on the forecast horizon and the presence of Stochastic Volatility (SV).

The results indicate that large predictor models, particularly those incorporating advanced priors without SV augmentation (e.g., LASSO large), outperform their SV-augmented counterparts. This finding contradicts the theoretical expectation that SV models should enhance accuracy by modeling time-varying uncertainty. A potential explanation lies in the increased noise and estimation challenges introduced by SV, particularly in high-dimensional settings. As a result, simpler models with broad predictor coverage provide a more robust and accurate forecasting framework.

From a macroeconomic forecasting perspective, our opinion aligns with contemporary research advocating for a balanced approach. In contexts where inflation dynamics are influenced by both domestic and global factors, relying solely on traditional, low-dimensional models may overlook critical information. At the same time, overly complex models risk losing generalizability due to overfitting.

In our view, policymakers should consider adopting shrinkage priors such as HS+ and DL in large predictor models while maintaining caution when integrating SV components. This is particularly true when forecasting under extreme economic conditions (e.g., the COVID-19 pandemic), where these priors demonstrate superior adaptability. Future research may explore hybrid modeling frameworks that dynamically switch between SV and non-SV models depending on the volatility regime.

While this study provides valuable insights into the forecasting performance of Bayesian shrinkage priors for Thai inflation, several areas for future improvement remain. One important direction involves exploring adaptive model selection strategies that dynamically switch between stochastic volatility and non-stochastic volatility frameworks based on changes in macroeconomic conditions. This approach would allow the model to better capture periods of heightened uncertainty while avoiding the noise and overfitting associated with SV during more stable phases.

Another promising avenue is to enhance the modeling framework by incorporating macro-financial variables such as global commodity prices, exchange rate fluctuations, and interest rate differentials. Including these external indicators may improve the model’s ability to capture the broader economic forces that drive inflation dynamics in an open economy like Thailand. Furthermore, time-varying parameter models with regime-switching capabilities could offer a more flexible structure that reflects the changing nature of inflation processes, particularly during major economic disruptions.

Future research could also benefit from employing Bayesian model averaging (BMA) to mitigate the risks associated with model selection uncertainty. By combining forecasts from multiple models rather than relying on a single specification, BMA may provide more stable and accurate predictions. Additionally, the integration of machine learning techniques such as gradient boosting or neural networks could complement traditional econometric models by capturing complex, non-linear relationships that standard Bayesian methods may overlook.

Finally, expanding the forecast evaluation window to include more recent data and emerging economic shocks would enhance the robustness and generalizability of the findings. Conducting a comparative analysis across different economies with varying inflation regimes could also provide a broader understanding of the general applicability of these Bayesian methods. These future improvements would not only refine the accuracy of inflation forecasting but also provide more practical insights for policymakers seeking to navigate uncertain economic environments.

\section{Conclusion}
\label{sec:conclusion}
This study examines the forecasting performance of various Bayesian shrinkage priors in predicting Thai inflation. The empirical results reveal an interesting counterparts: despite their theoretical advantages in capturing time-varying uncertainty, SV models consistently underperform relative to simpler models across multiple performance metrics, including Root Mean Squared Error (RMSE), Quantile-Weighted Continuous Ranked Probability Scores (qwCRPS), and Log Predictive Likelihood (LPL). This pattern is most evident in models incorporating a large number of predictors, where the introduction of stochastic volatility appears to increase estimation noise rather than improving forecast accuracy. Among the various models evaluated, the LASSO large model without SV consistently provides superior performance across both short- and long-term horizons, highlighting the efficacy of broad predictor coverage in enhancing predictive power.

An important observation emerging from the analysis is the asymmetry in forecast accuracy between left-tail and right-tail risks. While advanced shrinkage priors such as Horseshoe (HS), Horseshoe+ (HS+), and Dirichlet-Laplace (DL) effectively capture deflationary pressures, they struggle to accurately forecast inflationary spikes. This indicates that the models perform better when predicting lower-than-expected inflation outcomes rather than capturing unexpected surges in price levels. Furthermore, the study highlights a fundamental trade-off between model complexity and predictive performance. Although stochastic volatility is beneficial in low-dimensional models for short-term forecasts, its inclusion in high-dimensional models deteriorates accuracy across medium- and long-term horizons due to overfitting and increased estimation variability.

The broader implication of these findings is that a simpler modeling structure combined with advanced shrinkage priors provides more robust and reliable forecasts for Thai inflation. This insight is particularly relevant for policymakers who rely on inflation projections to guide monetary decisions. In volatile macroeconomic environments, such as those influenced by external shocks like the COVID-19 pandemic, models with a rich set of predictors but without unnecessary complexity provide more stable and accurate forecasts. This research contributes to the growing body of literature on macroeconomic forecasting by emphasizing the practical limitations of stochastic volatility models in high-dimensional settings and advocating for a balanced modeling approach that prioritizes both accuracy and interpretability.

In summary, our Horseshoe-prior analysis shows that under Thailand’s credible inflation-targeting framework (adopted in 2000 and refined in 2015), purely monetary and demand-side variables are nearly always driven to zero (low $\kappa$), leaving only genuine cost-push and structural factors as persistent inflation drivers. Across all rolling windows, categories such as Eggs \& Dairy, Seasoning \& Condiments, Non-Food Beverages, and Electricity/Fuel/Water consistently exhibit high $\kappa > 0.6$, reflecting their exposure to supply-chain shocks (e.g. the 2011 floods and the 2008 oil surge). Most notably, the Number of Alien Work Permits in Bangkok remains one of the very few predictors with $\kappa\approx 0.8$ even at the one-month horizon, underscoring how migrant-labor flows from neighboring countries immediately translate into wage-cost pressures that monetary policy cannot neutralize. These findings not only validate the global-local shrinkage prior sucha as Horseshoe’s ability to isolate true cost-push factors but also provide a clear, event-anchored narrative of why and how specific categories drive Thai inflation today.

\section*{Acknowledgments}
The author thanks Mae Fah Luang University for their support.

\section{Appendix}
\label{sec:appendix}
\subsection{Data}
\label{subsec:appendixDATA}
\begin{landscape}
\begin{table}[ht!]
\centering
\resizebox{1.4\textwidth}{!}{ 
\begin{tabular}{cll}
\hline
\textbf{no.} & \multicolumn{1}{c}{\textbf{Long-Acronyms}}                    & \multicolumn{1}{c}{\textbf{description}}                                                                                                                                            \\ \hline
1            & Consumer Price index                                          & The Thai CPI reflects the price movement of goods and services   across various categories, including:                                                                              \\
             & \multicolumn{1}{c}{}                                          & Food and Beverage, Housing and Utilities, Transportation and   Communication, Health and Education                                                                                  \\
2            & Production: Southern: Palm Oil: Crude                         & Crude palm oil (CPO) produced in southern Thailand, the   country's main palm oil-producing region.                                                                                 \\
3            & Motor Vehicle Production: Passenger Car: Less than 1500 cc    & Production of passenger cars with engine displacements of less   than 1500 cubic centimeters (cc) in Thailand.                                                                      \\
4            & Motor Vehicle Production: Passenger Car: 1501 to 1800 cc      & Production of passenger cars with engine displacements between   1501 and 1800 cubic centimeters (cc) in Thailand.                                                                  \\
5            & Motor Vehicle Production: Passenger Car: 1801 to 2000 cc      & production of passenger cars with engine displacements between   1801 and 2000 cubic centimeters (cc) in Thailand.                                                                  \\
6            & Motor Vehicle Production: Passenger Car: 2000 cc and Above    & production of passenger cars with engine displacements of 2000   cubic centimeters (cc) and above in Thailand                                                                       \\
7            & Motor.Vehicle.Production..Truck..10.Tons.and.Above            & production of trucks with a gross vehicle weight of 10 tons   and above in Thailand.                                                                                                \\
8            & Motor.Vehicle.Production..Truck..5.to.10.Tons                 & production of trucks with a gross vehicle weight between 5 and   10 tons in Thailand.                                                                                               \\
9            & Motor.Vehicle.Production..Truck..Less.Than.5.Tons             & production of trucks with a gross vehicle weight of less than   5 tons in Thailand.                                                                                                 \\
10           & Govt Revenue: Gross: Excise Dept: Others Tax (OT)             & gross revenue collected by the Excise Department from various   other taxes in Thailand.                                                                                            \\
11           & Govt Revenue: Gross: Excise Dept: OT: Perfume                 & gross revenue collected by the Excise Department from taxes on   perfume products in Thailand                                                                                       \\
12           & Govt Revenue: Gross: Excise Dept: Miscellaneous               & gross revenue collected by the Excise Department from various   miscellaneous taxes in Thailand.                                                                                    \\
13           & Govt Revenue: Gross: Customs Dept: Import Duties              & gross revenue collected by the Customs Department from import   duties in Thailand.                                                                                                 \\
14           & Govt Revenue: Gross: Customs Dept: Export Duties              & gross revenue collected by the Customs Department from export   duties in Thailand                                                                                                  \\
15           & Govt Revenue: Gross: OS: Other Government Sections            & gross revenue collected by various other government sections   in Thailand.                                                                                                         \\
16           & Govt Revenue: Gross: OS: State Enterprises                    & gross revenue collected by state enterprises in Thailand.                                                                                                                           \\
17           & Govt Revenue: Gross: Revenue Dept: Personal Income Tax        & gross revenue collected by the Revenue Department from   personal income taxes in Thailand.                                                                                         \\
18           & Govt Revenue: Gross: Revenue Dept: Corporate Income Tax       & gross revenue collected by the Revenue Department from   corporate income taxes in Thailand.                                                                                        \\
19           & Govt Revenue: Gross: Revenue Dept: Value Added Tax            & gross revenue collected by the Revenue Department from   value-added tax (VAT) in Thailand.                                                                                         \\
20           & Govt Revenue: Gross: Revenue Dept: Specific Sale Tax          & gross revenue collected by the Revenue Department from   specific sales taxes in Thailand.                                                                                          \\
21           & Govt Revenue: Gross: Revenue Dept: Stamp Duties               & gross revenue collected by the Revenue Department from stamp   duties in Thailand.                                                                                                  \\
22           & Govt Revenue: Gross: Revenue Dept: Others                     & gross revenue collected by the Revenue Department from various   other sources in Thailand.                                                                                         \\
23           & Govt Revenue: Gross: Excise Dept: Tobacco Tax                 & gross revenue collected by the Excise Department from tobacco   taxes in Thailand.                                                                                                  \\
24           & Govt Revenue: Gross: Excise Dept: Beer Tax                    & gross revenue collected   by the Excise Department from beer taxes in Thailand.                                                                                                     \\
25           & Govt Revenue: Gross: Excise Dept: Alcoholic Tax               & gross revenue collected by the Excise Department from   alcoholic taxes in Thailand. Includes                                                                                       \\
             &                                                               & all taxes levied on alcoholic beverages, such as beer, wine,   and spirits.                                                                                                         \\
26           & Govt Revenue: Gross: Excise Dept: Non Alcoholic Tax           & gross revenue collected by the Excise Department from   non-alcoholic taxes in Thailand. Includes all taxes levied on non-alcoholic   beverages and related products.               \\
27           & No of New Applicants: Bangkok                                 & number of new job applicants in Bangkok, Thailand. Includes   the total number of individuals who have applied for jobs in Bangkok.                                                 \\
28           & No of New Applicants: Other Regions                           & number of new job applicants in regions outside of Bangkok,   Thailand. Includes the total number of individuals who have applied for jobs   in various regions across the country. \\
29           & No of Job Vacancies: Female                                   & number of job vacancies available for female applicants in   Thailand. Includes the total number of job openings specifically for female   candidates.                              \\
30           & No of Job Vacancies: Non-Specific                             & number of job vacancies that are not specifically categorized   by gender or region in Thailand.                                                                                    \\
31           & No of Job Vacancies: Bangkok                                  & number of job vacancies available in Bangkok, Thailand.                                                                                                                             \\
32           & No of Job Vacancies: Male                                     & number of job vacancies available for male applicants in   Thailand. Includes the total number of job openings specifically for male   candidates.                                  \\
33           & No of Job Placement: Bangkok                                  & number of job placements in Bangkok, Thailand. Includes the   total number of individuals who have successfully secured jobs in various   sectors within the city.                  \\
34           & No of Aliens Work Permit: Illegal Immigrant                   & number of work permits issued to illegal immigrants in   Thailand.                                                                                                                  \\
35           & No of Aliens Work Permit: Bangkok: Temporary                  & number of temporary work permits issued to foreign workers in   Bangkok, Thailand.                                                                                                  \\
36           & No of Aliens Work Permit: Bangkok: Illegal Immigrant          & number of work permits issued to illegal immigrants in   Bangkok, Thailand.                                                                                                         \\
37           & No of Aliens Work Permit: Other Regions: Investment Promotion & number of work permits issued to foreign workers in regions   outside of Bangkok, Thailand, under investment promotion schemes.                                                     \\
38           & No of Aliens Work Permit: Other Regions: Temporary            & number of temporary work permits issued to foreign workers in   regions outside of Bangkok, Thailand.                                                                               \\
39           & BOT: No of Job Placements                                     & number of job placements facilitated by the Bank of Thailand   (BOT).                                                                                                               \\
40           & BOT: No of Overseas Thai Worker: Others                       & number of Thai workers employed overseas in various countries,   excluding specific countries like Israel, Japan, Malaysia, Singapore, and   Taiwan.                                \\
41           & BOT: No of Overseas Thai Worker: Israel                       & number of Thai workers employed in Israel.                                                                                                                                          \\
42           & BOT: No of Overseas Thai Worker: Japan                        & number of Thai workers employed in Japan.                                                                                                                                           \\
43           & BOT: No of Overseas Thai Worker: Malaysia                     & number of Thai workers employed in Malaysia.                                                                                                                                        \\
44           & BOT: No of Overseas Thai Worker: Singapore                    & number of Thai workers employed in Singapore.                                                                                                                                       \\
45           & BOT: No of Overseas Thai Worker: Taiwan                       & number of Thai workers employed in Taiwan.                                                                                                                                          \\
46           & BOT: Inward Remittances                                       & inward remittances received in Thailand. Includes the total   amount of money transferred into Thailand from abroad.                                                                \\
47           & Motor Vehicle Sales: TMT: Passenger Car                       & sales of passenger cars in Thailand.                                                                                                                                                \\
48           & CPI: FB: CH: Eggs and Dairy Products                          & Consumer Price Index (CPI) for eggs and dairy products in   Thailand. Includes the price changes of eggs and dairy products.                                                        \\
49           & CPI: FB: CH: Seasoning and Condiments (SC)                    & Consumer Price Index (CPI) for seasoning and condiments in   Thailand. Includes the price changes of various seasoning and condiment   products.                                    \\
50           & CPI: FB: CH: Non Alcoholic Beverages                          & Consumer Price Index (CPI) for non-alcoholic beverages in   Thailand. Includes the price changes of various non-alcoholic beverages.                                                \\
51           & CPI: HF: Electricity, Fuel, Water Supply                      & Consumer Price Index (CPI) for electricity, fuel, and water   supply in Thailand. Includes the price changes of these essential utilities.                                          \\
52           & CPI: HF: Textile of House Furnishing                          & Consumer Price Index (CPI) for textiles used in house   furnishings in Thailand. Includes the price changes of various textile   products used for home furnishings.                \\
53           & CPI: HF: Cleaning Supplies                                    & Consumer Price Index (CPI) for cleaning supplies in Thailand.   Includes the price changes of various cleaning products.                                                            \\
54           & CPI: Transport \& Communication (TC)                          & Consumer Price Index (CPI) for transportation and   communication services in Thailand. Includes the price changes of various   transportation and communication services.          \\
55           & CPI: Tobacco \& Alcoholic Beverages (TA)                      & Consumer Price Index (CPI) for tobacco and alcoholic beverages   in Thailand. Includes the price changes of various tobacco and alcoholic   beverage products.                      \\
56           & CPI: Non Food \& Beverages (NFB)                              & Consumer Price Index (CPI) for non-food and beverages in   Thailand. Includes the price changes of various non-food and beverage items.                                             \\ \hline
\end{tabular}}
\caption{Monthly Thai macroeconomic data used for small, moderate and large-sized model setup. All series are transformed into YoY growth rate. The first series is Thai consumer price index and the first 20 series are included in moderate-sized model setup, and all 56 are for large model setup.}
\label{tab:appenDATA}
\end{table}
\end{landscape}

\subsection{Additional Results on Driver of Thai inflation across forecasting horizons}
\label{subsec:appendixAddResultsDriver}

\clearpage
\section{Acknowledgements}
The authors declare that they have no known competing financial or non-financial interests that could have influenced the research, authorship, or publication of this article. Additionally the authors have no conflicts of interest to disclose.

\clearpage
\bibliographystyle{apalike} 
\bibliography{ref}
\end{document}